\documentclass[english,showpacs,twocolumn]{revtex4-1}

\usepackage{mathptmx}
\usepackage[T1]{fontenc}
\usepackage[latin9]{inputenc}
\usepackage{color}
\usepackage{amsmath}
\usepackage{amssymb}
\usepackage{hyperref}
\usepackage{graphics}
\usepackage{graphicx}
\usepackage{epsfig}
\usepackage{bm}
\usepackage{epstopdf}
\usepackage{mathtools}
\usepackage{soul}
\usepackage{subfigure}   

\usepackage{xmpmulti}

\def\beq{\begin{equation}}
\def\eeq{\end{equation}}

\def\reff#1{(\ref{#1})}

\def\Up{U_\mathrm{p}}

\def\Wcmcm{\mbox{\rm Wcm$^{-2}$}}

\def\omegaMie{\omega_\mathrm{M}}
\def\omegaM{\omega_\mathrm{M}}

\def\vekt#1{\bm{#1}}

\def\v0{v_0}
\def\I0{I_0}

\usepackage{babel}

\begin{document}

\title{
Anharmonic resonance absorption of short laser pulses in clusters: A molecular dynamics simulation study
}

\author{S. S. Mahalik and M. Kundu}
\affiliation{Institute for Plasma Research, HBNI, Bhat, Gandhinagar - 382 428, Gujarat, India}
\date{\today}
\begin{abstract}
%
Linear resonance (LR) absorption of an intense 800~nm laser light in a 
nano-cluster requires a long laser pulse > 100~fs 
when Mie-plasma frequency ($\omegaMie$) of 
electrons in the expanding cluster matches the 
laser frequency~($\omega$). 
For a short duration of the pulse 
the condition for LR is not satisfied. 
In this case, it was shown 
by a model and particle-in-cell (PIC) simulations 
[Phys. Rev. Lett. 96, 123401 (2006)] that
electrons absorb laser energy by 
anharmonic resonance (AHR) when the position-dependent frequency 
$\Omega[r(t)]$ of an electron in the self-consistent 
anharmonic potential of the cluster satisfies $\Omega[r(t)]=\omega$. 
However, AHR remains to be a debate and still obscure in multi-particle plasma simulations. 
Here, we  identify AHR mechanism in a laser driven cluster using
molecular dynamics (MD) simulations. 
By analyzing the trajectory of each 
MD electron and extracting its $\Omega[r(t)]$ in 
the self-generated anharmonic plasma potential it is found 
that electron is outer ionized {\em only} when AHR is met.
An anharmonic oscillator 
model, introduced here, brings out most of the features of MD electrons 
while passing the AHR. Thus, we not only bridge the gap between PIC 
simulations, analytical models and MD calculations for the first time but also 
unequivocally prove that AHR processes is a universal dominant 
collisionless mechanism of absorption in the short pulse regime or in 
the early time of longer pulses in clusters. 
\end{abstract}
\maketitle

\section{Introduction}\label{sec1}
Laser-driven atomic clusters absorb large 
fraction of laser energy compared to traditional 
solid and gas targets. 
Solid like overdense plasma density of a 
cluster and its smaller size (of a few nanometer) than the 
wavelength of 800 nm laser pulse (typically used in experiments) allow 
full penetration of laser field without its attenuation, contrary to 
micron-sized solids, leading to nearly 90\%
laser absorption in clusters \cite{Ditmire_PRL78}. 

After the irradiation of a cluster by a laser pulse of 
intensity > $10^{14}\,\Wcmcm$, individual atoms in the cluster are 
ionized (called inner ionization) and a nano-plasma is created. 
Subsequent interactions lead to absorption of laser
energy by electrons and removal of those energetic electrons
from the time-dependent cluster potential (called outer ionization). As 
an electron crosses the cluster boundary or leaves the cluster 
completely, a local electrostatic (ES) field due to charge non-neutrality 
develops. This ES field together with the laser field may lead to
further inner ionization (called ionization ignition \cite{RosePetruck}) and creation of 
higher charge states which are forbidden by the laser 
field alone. Higher ionic charge states from rare gas clusters 
at a lower intensity than required for isolated atoms confirm the role 
of this ionization ignition. 
The electrostatic energy 
stored in the non-neutral nano-plasma (after some electrons 
are removed by laser) gets converted to ion kinetic energies through 
ion-ion Coulomb repulsion resulting MeV ions 
\cite{Ditmire_PRL78,Ditmire_Nature386,Ditmire_PRL78_2732,Kumarappan_PRL87,Lezius,Fukuda,
Kumarappan2001,Krishnamurthy,Kumarappan2002,Ditmire_PRA57} in experiments. KeV electrons \cite{Kumarappan2002,Ditmire_PRA57,Springate_PRA61,Springate_PRA68,Shao_PRL77,Chen_POP_9,Kumarappan2003}, x-rays \cite{Jha_2005,Jha_2006,Chen_PRL104,McPherson_Nature370} 
and MeV neutrals \cite{Rajeev_Nature} in experiments 
are the consequence of efficient coupling of laser energy 
with cluster.

Clearly, without the efficient electron acceleration, the subsequent 
ion acceleration, x-ray generation and neutral atom acceleration can 
not happen. 
Therefore, investigation of underlying physical process of coupling of 
laser energy with electrons is very important. 
For infrared lasers 
(800 nm wavelength or above) with intensities $\I0 >10^{16}\,\Wcmcm$, 
 collisional process of absorption through electron-ion collision can 
be neglected since it scales as $\sim I_0^{-3/2}$.
However, collisionless process of absorption continues irrespective
of the laser intensity and the wavelength as long as the plasma 
is overdense.  

Among the various collisionless processes, linear resonance (LR) occurs 
when Mie-plasma frequency $\omegaM(t)$ of the Coulomb expanding cluster 
meets the laser frequency $\omega$. In experiments, it is achieved by 
a pump-probe technique \cite{Zweiback,Saalmann_JPB39,Fennel_RMP} where a pump pulse first ionizes the cluster, 
$\omegaM(t)$ rises above $\omega$, 
subsequent outer ionization of electrons leads to cluster expansion 
causing decrease of $\omegaM(t)$ towards $\omega$, and after a suitable 
delay (typically > 100 fs) a probe pulse hits the expanding cluster 
to meet the LR condition $\omegaM (t) \approx \omega$ \cite{Ditmire_PRA53}. However, to make 
this LR to happen at a later time, we need to remove as 
many electrons by a laser from the cluster potential in an early time. 
Otherwise Coulomb expansion is not possible. 
Therefore, one must know 
how those electrons absorb energy in an early time of a long pulse laser. 
For a short pulse, ionic background can not expand sufficiently,
$\omegaM(t)$ does not fall upto $\omega$ to meet LR; 
{\em but} energy absorption by electrons still persists. 
Therefore, we concentrate on the mechanism of energy absorption in the 
short pulse regime which is also applicable for the early duration of a 
longer pulse much before the LR can happen.

The role of $\vec{v} \times \vec{B}$ heating \citep{Kruer}
as a collisionless mechanism is discriminated here by restricting the 
laser intensity below $10^{17}\,\Wcmcm$ where $B$ field of the laser 
is negligible. 
One may surmise the ``Brunel effect'' or the ``vacuum heating'' 
as a probable collisionless process  
in the early time of interaction when plasma boundary is sharp \cite{Brunel}. 
Firstly, the ``vacuum'' as mentioned by Brunel may not be a real vacuum.
Electrostatic field exists in the target vicinity (as we shall show here) 
that plays a 
crucial role for an electron's dynamics before it is liberated from
the target or directed into the target. 
It can not gain a net energy, 
unless there is any nonlinear interaction through the nonlinear 
space-charge field within the target and/or in the target vicinity. 
Therefore, the tautological name ``vacuum heating'' is improper.  
According to Brunel's original proposition \cite{Brunel}, when an intense laser 
pulse strikes a sharply bounded overdense plasma;
electrons are dragged in to the vacuum and then due to the 
laser field reversal, in the next half-cycle of the pulse, electrons are 
pushed back inside the target with a velocity on the order of 
the ponderomotive velocity $v_0= e E_0/m \omega$. The {\em crucial}
assumption in this model is that electrons experience {\em no net} 
field while they return to the target. As a consequence
Brunel's electron flow becomes laminar, meaning that 
their trajectories do not cross each 
other within a laser period irrespective of the laser 
intensity. 
We point out that, since different electrons originate from different parts of the target
they experience different electrostatic fields and originate with different
initial phases. When driven by a laser, their trajectory 
crossing is unavoidable at a later time.
Detail analysis showing deficiency in Brunel's ``vacuum heating'' is given in Ref.\cite{Mulser2012}.
Brunel electrons, upon 
returning to the target, experience a field free 
region due to complete cancellation of induced electrostatic field by 
the laser field. 
Thus, the velocities acquired during their traversal in 
the vacuum are fully retained, they do not have chance to give energy back 
(even partly) to the electromagnetic field. In the context of 
laser-cluster interaction, induced electrostatic fields can not be fully 
compensated by the laser field (induced field may exceed the 
laser field) and cluster interior is rarely field free 
(as shown in this work) during 
the laser interaction. Otherwise, ionization ignition \cite{RosePetruck} can not happen 
and higher charge states \cite{Bauer2003,Ishikawa,Bauer2004,Megi} 
of ions can not be created. In this sense, Brunel effect is 
incomplete, warrant a re-look into the problem and search for an
appropriate mechanism behind the laser absorption.  

On the other hand, let us suppose that there is an anharmonic 
potential created at the target front (or in the target interior) 
due to the laser interaction. 
Such a potential is inevitably formed (for any finite size target) 
at the ion-vacuum boundary (where laser interacts first) 
due to $\sim 1/r$ fall of the potential which may be asymmetric. 
Anharmonicity in the potential also appears due to local charge 
non-uniformity (via ionization, concentrated electron cloud etc.)
in a laser driven plasma.
The frequency $\Omega$ of an electron in such a potential is 
dependent on its position $r$. 
When driven 
by a laser field, its $r$ changes with time which makes $\Omega(r)$ 
time dependent, i.e., $\Omega[r(t)]$.
An initially bound electron, starting from some location in the overdense 
plasma potential, while becoming free must experience the $\sim 1/r$  
Coulomb tail of the potential and the corresponding 
$\Omega[r(t)]$ of the electron must meet $\omega$ 
while trying to come out of the potential. 
This dynamical resonance 
- the anharmonic resonance (AHR) - occurring in an anharmonic potential
was studied before using a model and three dimensional PIC 
simulations of laser driven clusters \cite{MKunduprl}. 
However, collisionless processes and AHR phenomenon remain to be a debate \cite{Geindre_PRL2010,Mulser2015,Mulser2012}. 
In numerical simulations it is often obscured due to many body nature 
of interaction, 
since it needs clear examination of 
individual electron trajectory, identification of corresponding 
$\Omega[r(t)]$ and a dynamical mapping of $\Omega[r(t)]$ on to $\omega$.
To prove AHR for a laser driven cluster 
a three dimensional MD simulation code with soft-core 
Coulomb interactions among charge particles
has been developed. By following the 
trajectory of each MD electron and identifying its time-dependent 
frequency $\Omega[r(t)]$ in the self-generated anharmonic plasma potential 
it is found that electron leaves the potential and becomes free only 
when AHR condition $\Omega[r(t)]=\omega$ is met. 
Thus, for the first time, our MD simulation clearly identifies AHR process in the laser cluster interaction. 
We further introduce a non-linear oscillator model that brings out 
most of the features of MD electrons while passing the AHR. 
Thus, we bridge the gap between PIC simulations, analytical models and MD calculations. 

Atomic units (i.e., $m_e = -e =1, 4\pi\epsilon_0 = 1, \hbar = 1$) are used throughout this work unless specified explicitly.
We consider a single deuterium 
cluster of radius $R=2.05$~nm, Wigner-seitz radius {\textcolor{blue}{$r_w = R/N^{1/3}\approx 0.17$~nm}}
and number of atoms $N=1791$. It is irradiated by 800~nm wavelength 
laser pulses of various intensity giving density 
$\rho \approx 27.3\rho_c$ and $\omega_p^2/3\omega^2 = \omegaM^2/\omega^2 \approx 9.24$; where $\rho_c = \omega^2/4\pi$ is the critical density at
800~nm and $\omega_p$ is the plasma frequency. These parameters of the 
cluster are kept unchanged throughout this work.

Section~\ref{sec2} illustrates AHR by a simple model of a cluster while 
Sec.\ref{sec3} proves the hypothesis of AHR by detailed MD simulations. Summary and 
conclusion are given in Sec.\ref{sec4}.

\section{Model for anharmonic resonance absorption}\label{sec2}
Before studying the laser-cluster interaction by MD simulations, we show 
here various features of AHR by a model of a cluster which 
will provide an easy interpretation of 
MD results in Sec.\ref{sec3}. In the model, cluster is assumed to be pre-ionized and 
consists of homogeneously charged spheres of massive ions and much 
lighter electrons of equal radii $R_i = R_e = R$. 
When their centers coincide plasma becomes charge neutral. 
The motion of ions can be neglected for short laser pulses < 50 fs 
and non-relativistic laser intensities $<10^{18}\,\Wcmcm$ as 
considered in this work. Thus ion sphere provides a sharp boundary 
with zero density gradient scale-length at the vacuum plasma boundary. 

The equation of motion (EOM) of the electron sphere in a linearly polarized 
laser field along $x$-direction reads 
\begin{equation} \label{eq:ofmotion}
\frac{d^2\vec{r}}{dt^2}+\frac{\vec {r}}{r}g(r) = \hat{x} (q_e/m_e) E_l(t) 
\end{equation}
where $\vec{r} = \vec{x}/R$ and $r = \left|\vec {r}\right|$. 
The electrostatic restoring field 
\begin{equation} \label{eq:restoringforce}
g(r) = \omegaM^2 R \times \begin{cases}
r &\text{if $0\leq r \leq 1$}\\
{1}/{r^2} &\text{if $r \geq 1$,}
\end{cases}
\end{equation}
can be derived by Gauss's law. 
It shows that as long as the excursion $r$ of the
center of the electron sphere remains inside the ion sphere, it 
experience a harmonic oscillation with a constant eigen-frequency 
$\omegaMie$. Crossing the boundary of the ion sphere, it begins to experience 
the Coulomb force and its motion becomes anharmonic with gradual reduction in the
eigen-frequency for increasing excursion from the center of the ion 
sphere. 
The nonlinear resorting field \eqref{eq:restoringforce} is 
simpler than used earlier \cite{MKundupra2006,Popruzhenko2008}. Nevertheless,  
it exhibits all features AHR phenomena elegantly, e.g., 
prompt generation of electrons within a time much shorter than a 
laser period, crossing of electron trajectories and 
subsequent non-laminar electron flow \cite{Mulser2015}. 
It neglects the interaction of diffuse 
boundary of the electron sphere with the sharp boundary of the ion sphere,
{\em but} allows us to calculate $\Omega[r]$ of the electron sphere analytically for an arbitrary excursion 
which is not possible with the $g(r)$ in Refs.\cite{MKundupra2006,Popruzhenko2008}. 

The cluster size being much smaller 
than the laser wavelength $\lambda=800$~nm as considered in this work,
the dipole approximation for the laser vector potential
$A(z,t) = A(t)\exp(-i 2\pi z/\lambda) \approx A(t)$ is assumed. Thus, 
the effect of propagation of light (directed in $z$) is disregarded.
We take $A(t) = (E_0/\omega)\sin ^2 (\omega t/2n) \cos (\omega t)$ 
for $0<t<n T$; where $n$ is the number of laser period $T$, $n T$ is the 
total pulse duration and $E_0$ is the field strength corresponding to 
an intensity $I_0 = E_0^2$. 
The driving field ${E_l} (t)=-d{A} /dt$ reads
\begin{equation} \label{eq:laserfield}
{E_l} (t) = (E_0/\omega) \begin{cases}
\sum \limits_{i=1}^{3}c_i\omega_i\sin(\omega_i t) &\text{if \, $0 < t < nT$}\\
0 &\text{otherwise};
\end{cases}
\end{equation}
where $c_1=1/2, c_2=c_3= -1/4, \omega_1 = \omega, 
\omega_2 = (1+1/n)\omega$, and $\omega_3 = (1-1/n)\omega$. 
Eq.\eqref{eq:laserfield} 
leads to the correct dynamics of a free electron \cite{MKundupra2012}
even for very short sub-cycle pulses in contrast to the
often used $\sin^2$-pulses \citep{Mulserprl, Mulserpra, Saalmann2003, Saalmann_JPB39}. 
   
\subsection{Effective frequency of the electron sphere}\label{sec2a}
The electrostatic potential corresponding to Eq.\eqref{eq:restoringforce} reads
\begin{equation} \label{eq:potential}
\phi (r) = \omegaM^2 R^2 \times \begin{cases}
{3}/{2}-{r^2}/{2} &\text{if \, $0\leq r \leq 1$}\\
{1}/{r} &\text{if \, $r \geq 1$}.
\end{cases}
\end{equation}
In the absence of a driver, the eigen-period $T$ of oscillation of the 
electron sphere in the potential \eqref{eq:potential}
can be calculated from
\begin{equation} \label{eq:timeperiod}
T = \frac{4 R}{\sqrt{2}} \int_{0}^{r_{m}} \frac{dr}{\sqrt{\Phi(r_m) - \Phi (r)}}.
\end{equation}
Here $\Phi(r_m) = q_e \phi(r_{m})$ is the potential energy stored in the oscillator at an initial distance $r = r_{m}$ from where it is left freely in the potential at a time $t=0$.
For $0\le r_{m}\le 1$,
Eq.~\eqref{eq:timeperiod} yields a constant  
$T = {2 \pi}/{\omegaM}$ 
and the effective frequency of oscillation of the electronic sphere as 
\begin{equation} \label{freqeqn}
\Omega(r) = {2 \pi}/{T} = \omegaM.
\end{equation}
%
When $r_{m} > 1$, we write $T = T_1 + T_2$ with 
{\textcolor{blue}{
$T_1/4$ as the time required for $r=1$ to $r=0$ and 
$T_2/4 = \frac{R}{\sqrt{2}} \int_{r_{m}}^{1} \frac{dr}{\sqrt{\Phi(r_{m}) - \Phi (r)}}$ is the time required for $r= r_m$ to $r=1$ which give}} 
\begin{eqnarray}\label{Tperiod}
\nonumber
{\textcolor{blue}{
{\mathclap{
\!\! T_1 \!=\! \frac{4}{\omegaM}\!\! \Big[\!\sin^{-1}\!\!\left(\!\!{\sqrt{r_m/(3 r_m - 2)}}\right)\Big],}} }} \\ 
{\mathclap{
\!\! T_2 \!=\! \frac{(2 r_m)^{3/2}}{\omegaM}\!\! \Big[\!\sin^{-1}\!\!\left(\!\!{\sqrt{(r_m-1)/r_m}}\right)\! +\! \sqrt{\!\!(r_{m} -1)/{r_{m}^2}}\Big].}}
\end{eqnarray}
{\textcolor{blue}{$T_1$ is obtained from the harmonic solution 
$r_\mathrm{in}\!=\!\!\! \sqrt{R^2\!+\! v_R^2/\omegaM^2} \sin(\omegaM(t\!\! - \!\!T_2/4)\!\! +\!\!  
\arctan({\omegaM R/v_R}))$ inside the cluster satisfied by the 
electron that enters the surface of the cluster with the velocity 
$v_R=-\omegaM R\sqrt{2(r_m-1)/r_m}$ (obtained from 
the energy conservation) at $t=T_2/4$. If the 
electron starts at the surface of the cluster, i.e., at $r_m=1$, we get $T_1 = 2\pi/\omegaM$, $T_2 = 0$ and recover Eq.\reff{freqeqn} with $T=T_1$.}}
The effective frequency $\Omega(r) = 2\pi/(T_1 + T_2)$ now depends on 
the excursion amplitude $r=r_{m}$, since the electron 
sphere interacts with the nonlinear part of the restoring field. In a 
laser field excursion changes with time.
Thus $\Omega[r(t)]$ depends on $t$ when $r(t)>1$.
In more realistic MD simulations of clusters, 
there is no pre-defined 
potential (as Eq.\eqref{eq:potential}) in which electrons oscillate. 
Therefore, finding $\Omega(r)$ analytically is not possible in MD. 
From Eq.~\reff{eq:ofmotion} we formally write (in analogy with a 
harmonic oscillator) the square of $\Omega(r)$ as the 
ratio of restoring field to the excursion of the electron sphere \cite{MKunduprl}
\begin{equation} \label{EffFreqSq}
\Omega^2 [r(t)] = \frac{g[r(t)]}{r(t)} = \frac{\text{restoring field}}{\text{excursion}}.
\end{equation}
Note that for a harmonic oscillator above relation yields the correct 
eigen frequency $\Omega_h$ with $g(r) = \Omega_h^2 r$.

\begin{figure}
\includegraphics[width=0.95\linewidth]{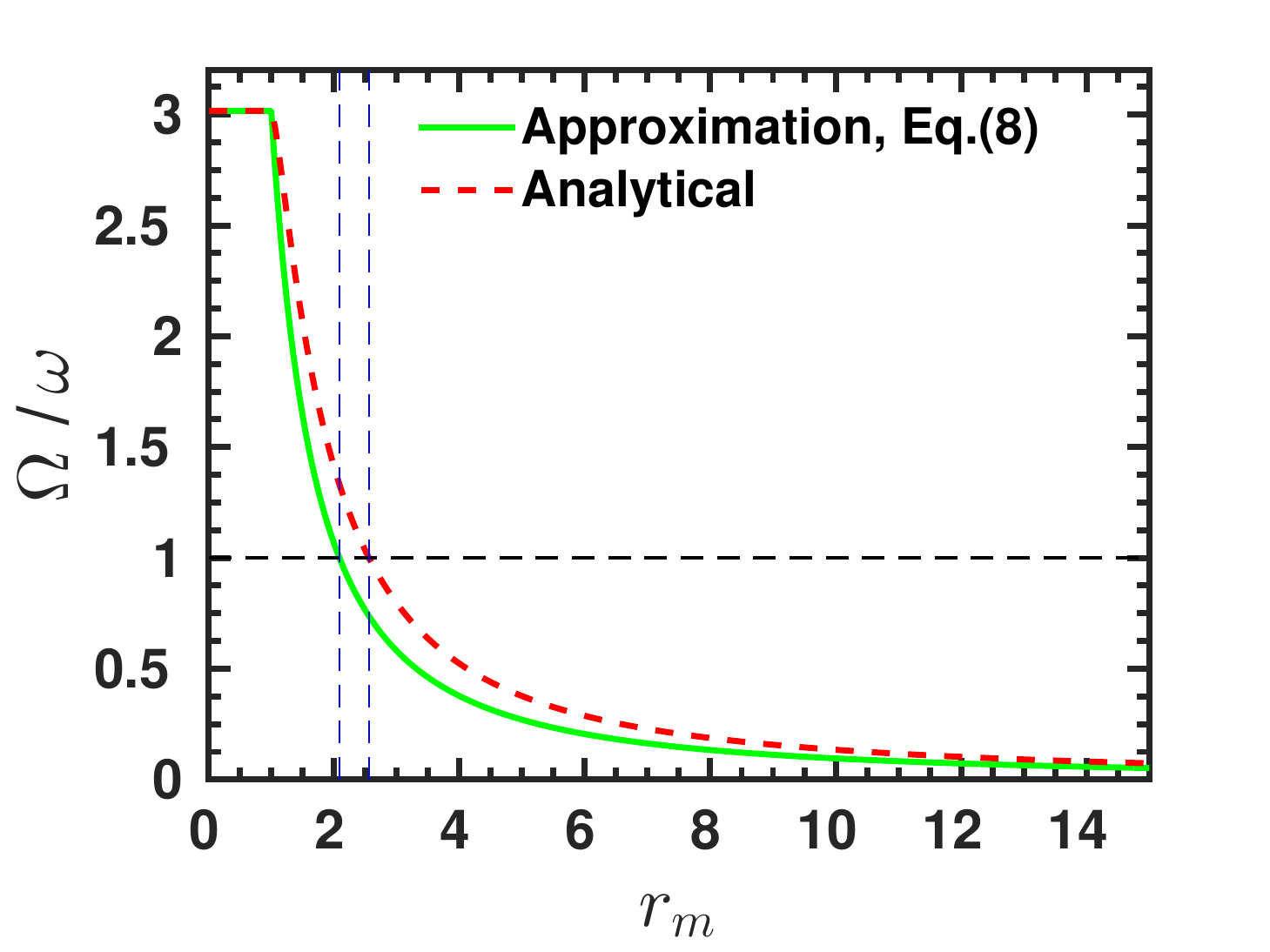}
\caption{Normalized effective frequency $\Omega/\omega$ of the electron sphere versus its excursion amplitude $r_{m}$ for a deuterium cluster of 
radius $R=2.05$~nm, Wigner-seitz radius $r_w \approx 0.17$~nm
and number of atoms $N=1791$,  density 
$\rho \approx 27.3\rho_c$ and $\omegaM^2/\omega^2 \approx 9.24$; where $\rho_c = \omega^2/4\pi$ is the critical density at
$\lambda = 800$~nm. Numerical approximation using Eq.\reff{EffFreqSq} 
and the analytical result using {\textcolor{blue}{Eqns.\reff{freqeqn}-\reff{Tperiod}}} are comparable.
{\textcolor{blue}{Vertical dashed lines indicate AHR is expected near $r_m \approx 2$ according to Eq.\reff{EffFreqSq} and $r_m\approx 2.5$ according to Eq.\reff{Tperiod}.}}
}
\label{fig1}
\end{figure}


Analytical result of the normalized effective frequency 
$\Omega/\omega$ using {\textcolor{blue}{(Eqns.\reff{freqeqn}-\reff{Tperiod})}} as a function of 
excursion ($r_m$) of the electron sphere in the potential 
(\ref{eq:potential}) is plotted in Fig.~\ref{fig1}. Numerical solution 
of Eq.~\reff{eq:ofmotion} (without the laser field) gives $r(t)$ and 
corresponding $\Omega/\omega$ from Eq.~\reff{EffFreqSq}. This numerical 
approximation is also plotted in Fig.~\ref{fig1} which
matches reasonably well with the analytical 
$\Omega/\omega$. As the electron sphere moves away from harmonic region 
of the potential, $\Omega/\omega$ starts decreasing. At a distance of 
$r_m \approx 2$, the AHR condition $\Omega/\omega \approx 1$ is 
satisfied. If the laser field is strong 
enough to bring the electron sphere at a value of 
$r(t) = r_m \approx 2$, the electron sphere may gain significant 
energy from the laser field via such AHR. 
\begin{figure}
\includegraphics[width=0.94\linewidth]{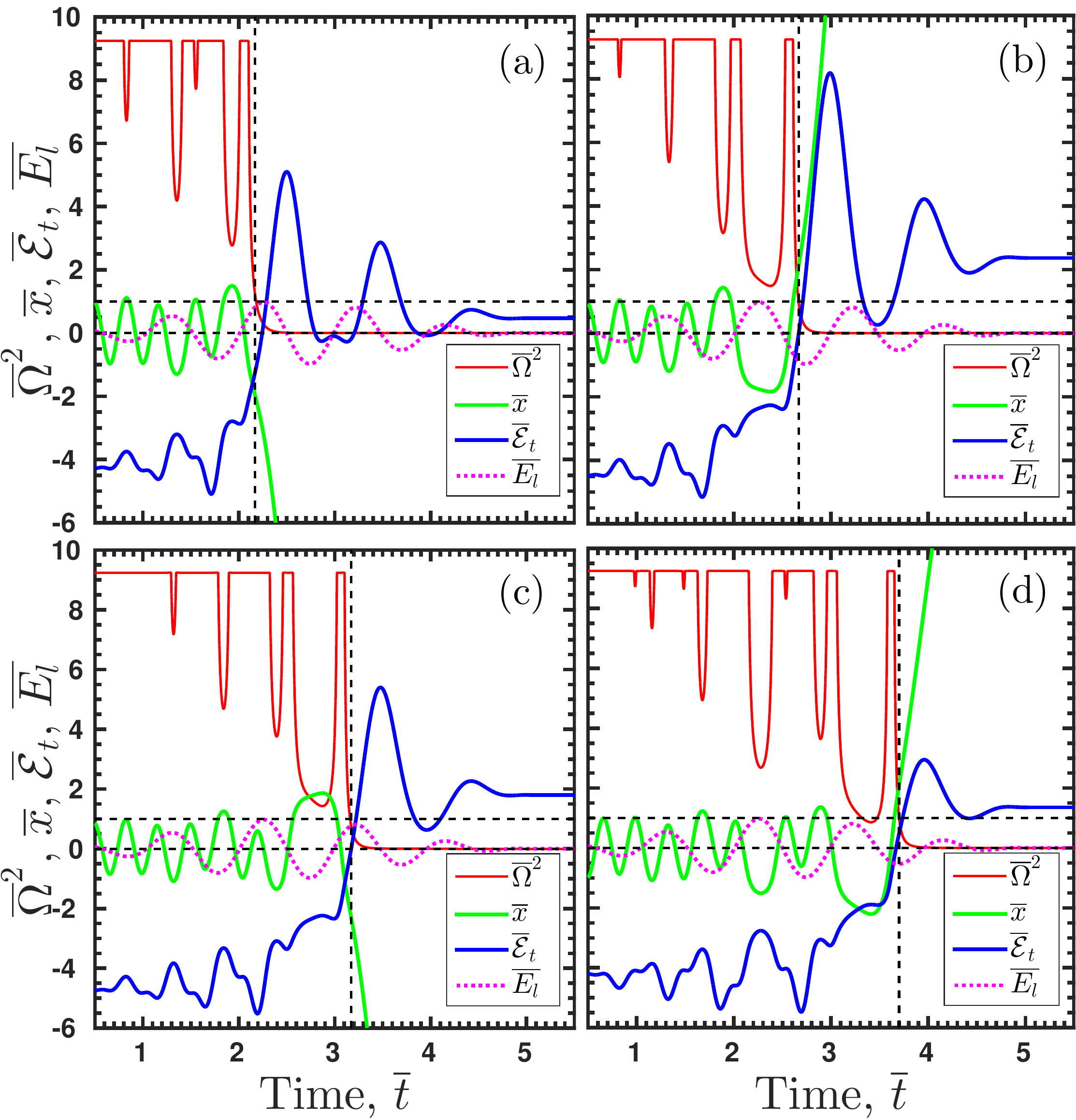}
\caption{Normalized value of the square of the 
effective frequency $\overline{\Omega}^2(r)$, 
excursion $\overline{x}$, 
total energy $\overline{\mathcal{E}}_t$ 
and the laser field $\overline{E}_l$ versus the normalized 
time $\overline{t}=t/T$ 
for four electron spheres undergoing AHR at successive times
$(a) \overline{t} = 2.2, (b) \overline{t} = 2.7, (c) \overline{t} = 3.2, (d) \overline{t} = 3.7$. 
The deuterium cluster of Fig.\ref{fig1} is irradiated by a 
$n = 5$-cycle pulse of peak intensity $5 \times 10^{15}\,\mathrm{W/cm^2}$. 
}
\label{fig2} 
\end{figure}

\subsection{Dynamics of the electron sphere in the laser field}\label{sec2b}
The dynamical behaviour of the electron sphere (for the above cluster) 
irradiated by a $n = 5$-cycle pulse (\ref{eq:laserfield})  of 
duration $nT = 13.5$~fs and peak 
intensity $5 \times 10^{15}\,\mathrm{W/cm^2}$ is now studied. 
Figure~\ref{fig2}(a-d) depicts the normalized value of the square of the 
frequency $\overline{\Omega}^2(r) = \Omega^2(r)/\omega^2$ 
using Eq.\reff{EffFreqSq}, excursion $\overline{x}=x/R$, 
total energy $\overline{\mathcal{E}}_t=\mathcal{E}_{t}/U_p$ 
and the laser field $\overline{E}_l = E_l/E_0$ versus the normalized 
time $\overline{t}=t/T$ for the electron sphere placed at four different 
initial locations $|\overline{x}|<1$ in the potential. 
Initially, spheres are bound 
with different energies (near $\overline{\mathcal{E}}_{t} \approx -4$) 
where they experience a constant frequency
$\Omega[r(0)] = \omegaM$. Up to $t/T\lesssim 2.0$, in Fig.\ref{fig2}(a), 
the laser field strength is not sufficient to liberate the electron sphere 
from the potential. In the oscillating laser field, at times 
$\overline{x}$ exceeds unity with a decrease in 
$\overline{\Omega}^2$ from $ \omegaM^2/\omega^2 \approx 9.24 $ and the 
corresponding increase in $\overline{\mathcal{E}}_t$. 
The decrease of 
$\overline{\Omega}^2[r(t)]$ being insufficient to meet the AHR condition
$\overline{\Omega}^2[r(t)]\approx 1$ (dashed horizontal line), particle can not absorb sufficient
energy to come out completely but pulled back towards the potential 
center (see the reversal of $\overline{x}$ and 
$\overline{\Omega}^2$ 
near $t/T\approx 1.9$) by the restoring force due to the ion sphere. 
As it reverses its direction and emerges on the other side of the 
potential with $\overline{x}<0$; the increasing laser field 
towards its peak value (i.e., $\overline{E}_l\sim 1$) 
after $t/T\approx 2.0$ 
helps $|\overline{x}|$ to exceed unity with a fast drop of 
$\overline{\Omega}^2$ for $t/T>2.1$. 
Around $t/T\approx 2.2$ (indicated by vertical dashed line), 
$\overline{\Omega}^2$ meets the AHR condition with 
$|\overline{x}|\approx 2$. The fact that AHR truly occurs at an 
excursion $|\overline{x}| \approx 2$ is 
in agreement with Fig.\ref{fig1}. It also justifies the 
robustness of the formal approximation (Eq.\reff{EffFreqSq}) for 
retrieving the effective frequency from the numerical model.
After the AHR (i.e., $t/T>2.2$) particle becomes completely free 
with $\overline{\Omega}^2[r(t)] \approx 0$ and final energy 
$\overline{\mathcal{E}}_{t} \ge 0$. 

Figures~\ref{fig2}(c-d) show electron spheres undergoing 
AHR at $t/T = 2.7, 3.2, 3.7$ respectively. 
Since initial positions are different, they have different initial
phases and experience different restoring fields and 
total fields even though same laser field acts on them. 
As a result they are emitted at different times from
the potential experiencing the AHR with different laser field strengths.
Electron spheres experiencing AHR at a higher driving field strength generally 
acquire higher energies after the pulse as in Figs.\ref{fig2}(a-c).
Some electrons may also exhibit [as in Fig.\ref{fig2}(d)] 
multiple AHR: they leave the cluster potential through AHR, return to 
the cluster interior by the laser field (or by the stronger restoring 
field than the laser field) and finally become free with a net positive 
energy via the AHR. 
%
%

\begin{figure}[]

\includegraphics[width=1.0\linewidth]{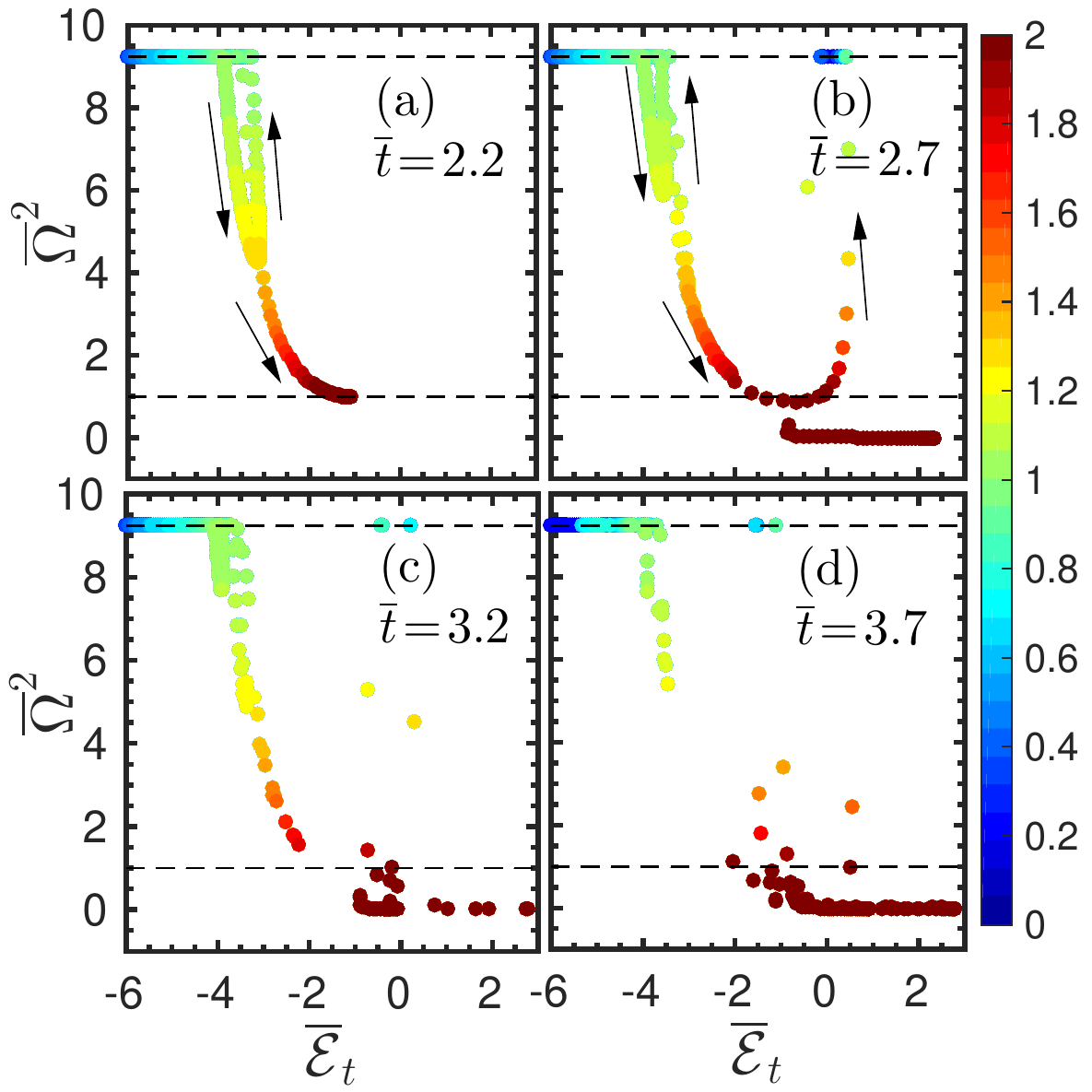}

\caption{(color online)
Snapshots of $N=1791$ non-interacting electron spheres 
in the (${{\overline{\mathcal{E}}}_{t}}$, $\overline{\Omega}^2$) plane
at times (a) $\overline{t} = 2.2$, (b) $\overline{t} = 2.7$, (c) $\overline{t} = 3.2$, (d) $\overline{t} = 3.7$ corresponding to Fig.\ref{fig2}. 
As the laser field strength increases with time, more and more electrons are drawn towards the line of AHR, i.e, dashed line at $\overline{\Omega}^2 \approx 1$.
The parameters of laser and cluster are same as in Fig.\ref{fig2}. 
}\label{fig3} 
\end{figure}
%

In this model electrons are frozen into a single sphere. 
Whereas, in reality, electrons in a cluster distinctly move even 
with a low intensity short laser pulse. 
Such a multi-electron system with all possible electrostatic 
interactions will be considered in detail by MD simulations in 
Sec.\ref{sec3}. To gain some more physical insight on the dynamics of 
electrons in a multi-electron cluster and the AHR through the above model, 
we consider $N=1791$ non-interacting electron spheres (mimicking the 
multi-electron system) placed uniformly inside the 
ion sphere. Each electron sphere mimics a real point size 
electron.

Figures~\ref{fig3}(a-d) show snapshots of all non-interacting 
electron spheres (each dot represents a sphere) in the energy versus 
effective frequency plane at times 
$\overline{t} = 2.2, 2.7, 3.2, 3.7$ corresponding to Figs.\ref{fig2}(a-d). 
Colors indicate their normalized positions.

In an early time $t/T = 2.2$ [in Fig.\ref{fig3}(a)] a large fraction of electrons are 
bound in the harmonic part of the potential with frequency 
$\overline{\Omega}^2 = \omegaM^2/\omega^2 = 9.24$ 
and excursion $r<1$ (dark blue to light blue). Some electrons first come out of the harmonic 
part and continue in the anharmonic part of the potential with a 
drop in $\overline{\Omega}^2$ (up to 4) and increasing excursion 
$1<r<1.3$ (light green to yellow). 
But the restoring field on them being higher than the
laser field they return to the cluster (see reversal of 
electrons with change of colors green-yellow-green in the U-shaped lobe) 
and become bound. At this time a few electrons (dark red to brown) are aligned towards the 
line of AHR (horizontal dashed line) with excursion $r\approx 2$ and frequency 
$\overline{\Omega}\approx 1$.
At a later time $t/T=2.7$ [in Figs.\ref{fig3}(b)], phase of the laser field is reversed 
with nearly the same strength. Apart from bound and returning electrons 
[as in Fig.\ref{fig3}(a)], it is clear that many electrons are now drawn towards the 
line of resonance, some are already free during this half laser cycles 
between $t/T=2.2-2.7$ 
due to change in the field. Beyond $t/T=2.5$ [in Figs.\ref{fig3}(b-d)] laser field 
strength becomes weaker than the restoring field on some quasi-free 
electrons. Those electrons (typically having low energies) are dragged 
inside the cluster (see their color changes from brown to yellow)
even though they were made free via the AHR earlier.

Thus a simple nonlinear oscillator model brings out most of the physics  
of AHR phenomena for a laser driven cluster in the temporal domain 
(Fig.\ref{fig2}) as well as in the energy versus frequency domain 
(Fig.\ref{fig3}). 
The identification of the effective dynamical frequency $\Omega[r(t)]$ 
of the driven oscillator in the numerical model and the liberation of 
particles from the cluster potential only when $\Omega$
matches the resonance condition ${\Omega}\approx \omega$, clearly 
justifies the robustness of the formal approximation in Eq.\reff{EffFreqSq}
and permits its application in the self-consistent MD simulations in 
Sec.\ref{sec3}.

\section{Anharmonic resonance absorption using molecular dynamics simulation}\label{sec3}
It is mentioned that in the above model electrons are frozen in a sphere 
which moves in a predefined attractive potential due to the ion sphere. 
In reality, the potential of the ionized cluster varies with the time 
to time redistribution of charges. In the initial time, when cluster is 
charge neutral, potential must start from a zero value. 
Electrons may also face repulsive potential due to concentrated electron 
cloud in some part of the cluster. In the model, the electron sphere 
either stays inside (0\% outer ionization) the cluster or completely 
goes out of the cluster (100\% outer ionization) for a given laser 
intensity. However, there is always a certain fraction of outer ionized 
electrons even at an intensity just above the inner ionization 
threshold. These shortcomings of the model can be addressed by MD 
simulation.

\subsection{Details of molecular dynamics simulation}\label{sec3a}
A three-dimensional MD simulation code is developed to study the 
interaction of laser light with cluster. Particular attention is given 
to the identification of AHR process using MD simulation. Cluster is 
assumed to be pre-ionized. This may be regarded as a situation where ionization has already taken 
place by a pump pulse and subsequent interaction of a probe pulse is studied.   
Electron-electron, ion-electron and ion-ion interactions through the Coulomb field are taken into account. 
Binary collisions among particles are neglected since we are interested 
in collisionless processes. The EOM of 
$i$-th particle in a laser field polarized in $x$ and propagating 
in $z$ (in the dipole approximation) reads
\begin{equation}\label{eom1}
m_i\displaystyle{\frac{d\vec{v_i}}{dt} = \vec{F_i}(r_i,v_i,t)} + \hat{x} q_i E_l(t) + q_i \vec{v_i}\times \hat{y} B_l(t), 
\end{equation}
where $\vec{F_{i}} = \sum\limits_{j=1, i\ne j}^{N_p}{q_i q_j}\vec{r_{ij}}/{{r_{ij}^3}}$ 
is the Coulomb force on $i$-th particle of charge $q_i$ due to all 
$j$-th particles each of charge $q_j$ in the system. $E_l(t)$ and $B_l(t)$ 
are the electric and magnetic part of the laser field.
Usually, $B_l(t)\approx E_l(t)/c \ll 1$ for intensities 
$< 10^{18}\,\Wcmcm$.
To avoid steep increase in the Coulomb force $\vec{F_i}$, for a small
separation $r_{ij} \rightarrow 0$, a smoothing parameter $r_0$ is added 
with ${r_{ij}}$. The modified Coulomb force on $i$-th particle and 
the corresponding potential at its location are 
\begin{equation}\label{MD_force}
{\vec F_{i}} = \sum_{j=1,i\ne j}^{N_p} \frac{q_i q_j {\vec{r_{ij}} } }{{(r_{ij}^2+r_0 ^2)^{3/2}}}, \,\,\,\,\,
\phi_{i} = \sum_{j=1, i\ne j}^{N_p} \frac{q_j}{{(r_{ij}^2+r_0 ^2)^{1/2}}}.
\end{equation}
This modification of the force allows a charge particle to pass through 
another charge particle in the same way as in the PIC simulation. Thus it helps to study collisionless energy absorption processes in plasmas, e.g., resonances. 

%
%
Equation \reff{eom1} is solved using the velocity verlet time integration 
scheme \cite{Verlet} with a uniform time step $\Delta t = 0.1$ a.u.. 
The code is validated by verifying the energy conservation of the system 
as well as identifying the electron plasma oscillation with desired 
Mie-plasma frequency $\omegaM$ for the spherical cluster plasma.
Although there are plenty of MD simulations for clusters \citep{RosePetruck,Ishikawa,LastJortner1999,LastJortner2000,LastJortner2004,Saalmann2003,Fomichev_pra,
Petrov2005_PRE,Petrov2005,Petrov2006,Petrov2007,G_Mishra2011,G_Mishra,Cheng,
Amol,Greschik&Kull,Batishchev,Bystryi,Arbeiter}, 
verification of this natural oscillation through MD simulation is rarely 
reported 
which is extremely important, particularly to study frequency dependent phenomena, e.g., anharmonic and harmonic resonance absorption in laser 
driven plasmas. 
{{Otherwise, resonance physics may be missing in simulations
and subsequent MD results could be misleading.
In fact, MD codes in Refs.\cite{Petrov2005_PRE,Petrov2005,Petrov2006,Petrov2007} could not find the signature of resonances in laser cluster interaction 
and the reason was unknown; while experiments \citep{Ditmire_PRA53,Doppner,Koller,Zamith}, theory and particle-in-cell simulation \citep{MKunduprl,MKundupra2006,Mulserprl,Mulserpra,Taguchi_PRl,Antonsen,Kostyukov} studies clearly indicated its importance.}} 

We note that the artificial free parameter $r_0$ in most of the earlier
works has been chosen 
by considering {\em only} the energy conservation point of view in the 
simulation.
%
In Refs.\citep{LastJortner1999,LastJortner2000,LastJortner2004}, $r_0 = 0.02$~nm for electron-electron interaction and $r_0 = 0.1$~nm for electron-ion interaction have been taken which do not violate the energy conservation in the case of xenon cluster. In Ref.~\citep{Fomichev_pra}, $r_0 = 0.15$~nm for argon and $r_0 = 0.12$~nm for xenon cluster were chosen such that the minimum of the electron-ion interaction potential agrees with the ionization potential of the neutral atom. MD codes in Refs.~\citep{Petrov2005_PRE,Petrov2005,Petrov2006,Petrov2007,G_Mishra2011,G_Mishra,Cheng} have reported similar kind of $r_0$ values as in Refs.~\citep{LastJortner1999,LastJortner2000,LastJortner2004}. Some of the authors have also chosen large $r_0$ on the order of cluster radius \citep{RosePetruck,Amol} and energy conservation is still obeyed. 

{{In our simulation, frequency of oscillation ($\omegaM$) of electrons is 
found to be
sensitive on the value of the $r_0$ while conservation of 
energy is obeyed even for larger values of $r_0$. But energy conservation 
{\em alone} can not grant correctness of particle dynamics in simulations. 
To get correct oscillation frequency ($\omegaM$) one can not choose $r_0$ 
arbitrarily and a law has to be enforced. 
For a very small separation $r_{ij}\ll r_0$, the space charge field on 
the $i$-th particle ${\vec E_i}^{sc} = {\vec F_i}/q_i$ has to be linear in $r_{ij}$ and
its slope has to be $\omegaM^2$, i.e., ${\vec E_i}^{sc} = {\vec F_i}/q_i 
= \omegaM^2 {\vec r}_{ij}$ in order to get correct plasma oscillations.
From Eq.\reff{MD_force}, we find that 
${\vec E_i}^{sc} = {\vec F_i}/q_i 
\approx \sum_j (q_j/r_0^3) {\vec r}_{ij} 
\approx (Q_0/r_0^3) {\vec r}_{ij}$ for $r_{ij}\ll r_0$; assuming 
$Q_0$ is the total uniformly distributed charge of type $j$ inside 
the sphere of radius $r_0$ where all $r_{ij}$ are nearly same for 
collective plasma oscillations. {\textcolor{blue}{ 
From the above two expressions of space-charge field ${\vec E_i}^{sc} = \omegaM^2 {\vec r}_{ij}$ 
and ${\vec E_i}^{sc} \approx (Q_0/r_0^3) {\vec r}_{ij}$ we get $\omegaM^2 = Q_0/r_0^3$. For uniform ionic charge density $\rho$ we write 
$\rho = Q/(4\pi/3)R^3 = Q_0/(4\pi/3)r_0^3 $ 
which gives $Q/R^3 = Q_0/r_0^3 = 4\pi\rho/3 = \omegaM^2$. 
Thus we get $Q_0/r_0^3 = N_0 Z/r_0^3 = Q/R^3 = N Z/R^3$ and
$r_0 = R (N_0/N)^{1/3}$; where $Z$ is the uniform charge state of ions 
in the cluster, $N_0, N$ are the number of ions inside 
the sphere of radius $r_0$ and $R$, {\textcolor{blue}{$Q_0=N_0 Z, Q = N Z$ 
are the total ionic charge in the sphere of radius $r_0$  and in the 
cluster respectively. At this point $N_0$ remains arbitrary. 
We note that $r_0$ should be as small as possible, but non-zero. 
For a non-zero ionic charge density there should be at least one ion 
(to provide the restoring force to an electron) 
in the sphere of radius $r_0$. 
}} 
Therefore, setting $N_0 = 1$, we find that $r_0 = R/N^{1/3}$ 
is the most legitimate choice \cite{Greschik&Kull} which is equal to the 
well known Wigner-Seitz radius $r_w = R/N^{1/3}$ for a given cluster that leads to correct Mie-plasma oscillation 
if the law of force is of Coulombic in nature.
}}

To prove that our MD code is capable of producing oscillation of electrons 
at the Mie-plasma frequency in the absence of a laser field, 
the deuterium cluster of radius 
$R = 2.05$~nm, and number of atoms $N = 1791$ (as in Sec.\ref{sec2}) is considered. 
To make homogeneously 
charged positive background in which electrons will oscillate, 
all $N_i$ ions ($N_i = N = 1791$) are uniformly distributed initially.  
It gives ionic charge density $\rho_i=0.007\,\mathrm{a.u.}$ 
and $\omegaM=\sqrt{4 \pi \rho_i/3} = 0.1735\,\mathrm{a.u.}$.  
A fewer number of $N_e$ electrons (forming a homogeneous 
sphere of radius $R/2$) is uniformly and symmetrically distributed about 
the center of the spherical ion background and the whole system is at 
rest. 
This may represent a situation when most of the electrons are removed 
from the cluster by a laser field, and the remaining cold electrons 
occupying the central region collectively oscillate with the 
frequency $\omegaM$. 
Electrons are now uniformly shifted (small perturbation) from the 
ionic background along $x$. The space charge field due to the local 
charge imbalance acts like a restoring field. Whether these 
electrons will oscillate at $\omegaM$ is determined by the homogeneity 
of the charge density $\rho_i$ and the linearity of the restoring field 
decided by the amount of perturbation. These are ensured by making the ion 
background homogeneous and keeping perturbation 
small so that electrons do not cross 
the cluster boundary. Under this condition one may write EOM 
of the center of mass of the electron cloud as
$\ddot{x} = \omegaM^2 x,$
giving $x = x_0 \cos(\omegaM t)$ with initial conditions 
$x(0) = x_0$, $\dot{x}(0) = 0$; and verify the MD simulation results.

The Fourier transform (FT) of the center of 
mass position $X_{cm}(t) =\sum_{1}^{N_e} m_e x_i(t) / N_e$ of 
MD electrons gives the collective oscillation frequency of the electron 
cloud which is plotted with the FT of the analytical solution 
$x = x_0 \cos(\omegaM t)$ in Fig.~\ref{fig4} after the normalization 
by $\omegaM$. An excellent match between the numerical (dashed-circle) 
and analytical (solid line) results 
confirms the collective oscillation of electrons at the Mie-plasma 
frequency by MD simulations.
{\textcolor{blue}{Extensive simulations have been performed for 
other values of $r_0=0.5 r_w, 2 r_w, 3 r_w, 4 r_w$ to check the 
effect of $r_0$ on the plasma oscillation dynamics. 
As time progresses, we find 5-20\% reduction in the 
amplitude of plasma oscillation with 
$5-10\%$ elongation in the plasma period 
compared to the case of $r_0=r_w$ and the desired analytical solution
shown in Fig.\ref{fig4}.
}}


\begin{figure}

\includegraphics[width = 0.85\linewidth]{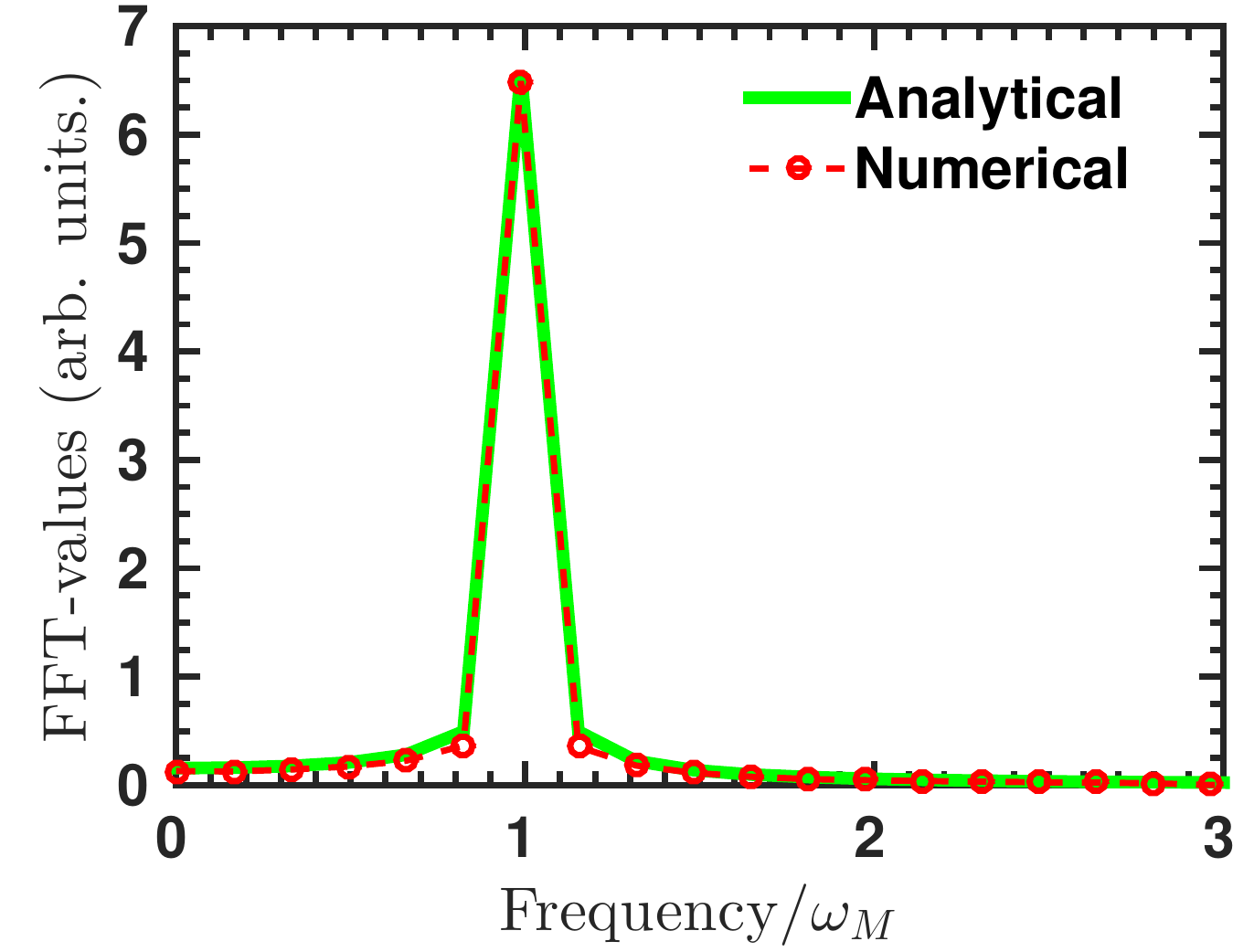}

\caption{(color online) FFT of the center of mass (CM) position co-ordinate of electrons for a deuterium cluster of 
radius $R=2.05$~nm, and number of atoms $N=1791$.  
Frequency is normalized by $\omegaM$. MD simulation result (numerical, dashed-circle) matches with the analytical result (solid line) for the entire range of frequency.
} 
\label{fig4}
\end{figure} 
\subsection{Laser energy absorption and outer ionization}
\label{sec3b}
The MD code is now used to study interaction 
of $n=5$-cycle laser pulse of duration $n T=13.5$ fs 
with the deuterium cluster as in Sec.\ref{sec2}.
Initially, cluster has equal number of uniformly 
distributed ions and electrons (i.e., $N_i=N_e=N$) so that 
it is macroscopically charge neutral. 
%

Figure~\ref{fig5} shows total absorbed energy per electron normalized by 
of $\Up$ and corresponding degree of outer ionization at the end of the 
laser pulse as the peak intensity is varied. 
Normalized absorption per electron [in Fig.\ref{fig5}(a)] attains a maximum between intensities
$5\times10^{15}-10^{16}\, \Wcmcm$. 
This nonlinear variation of absorbed 
energy with intensity is similar to that reported earlier using PIC 
simulations \cite{MKunduprl} of xenon clusters. 
The outer ionized fraction [in Fig.\ref{fig5}(b)] of electrons, on the other hand, 
increases gradually with the peak intensity 
and saturates at unity (\%100 outer ionization) at some higher 
intensity even for this short 5-cycle pulse. 
At an intensity $5\times10^{15}\,\Wcmcm$, 
it is inferred that 
almost $60\%$ electrons are outer ionized ($\overline{N} \approx 0.6$) 
which contribute to the total absorbed energy of $\approx 2000\, U_p$. 


\begin{figure}[]
\includegraphics[width = 0.85\linewidth]{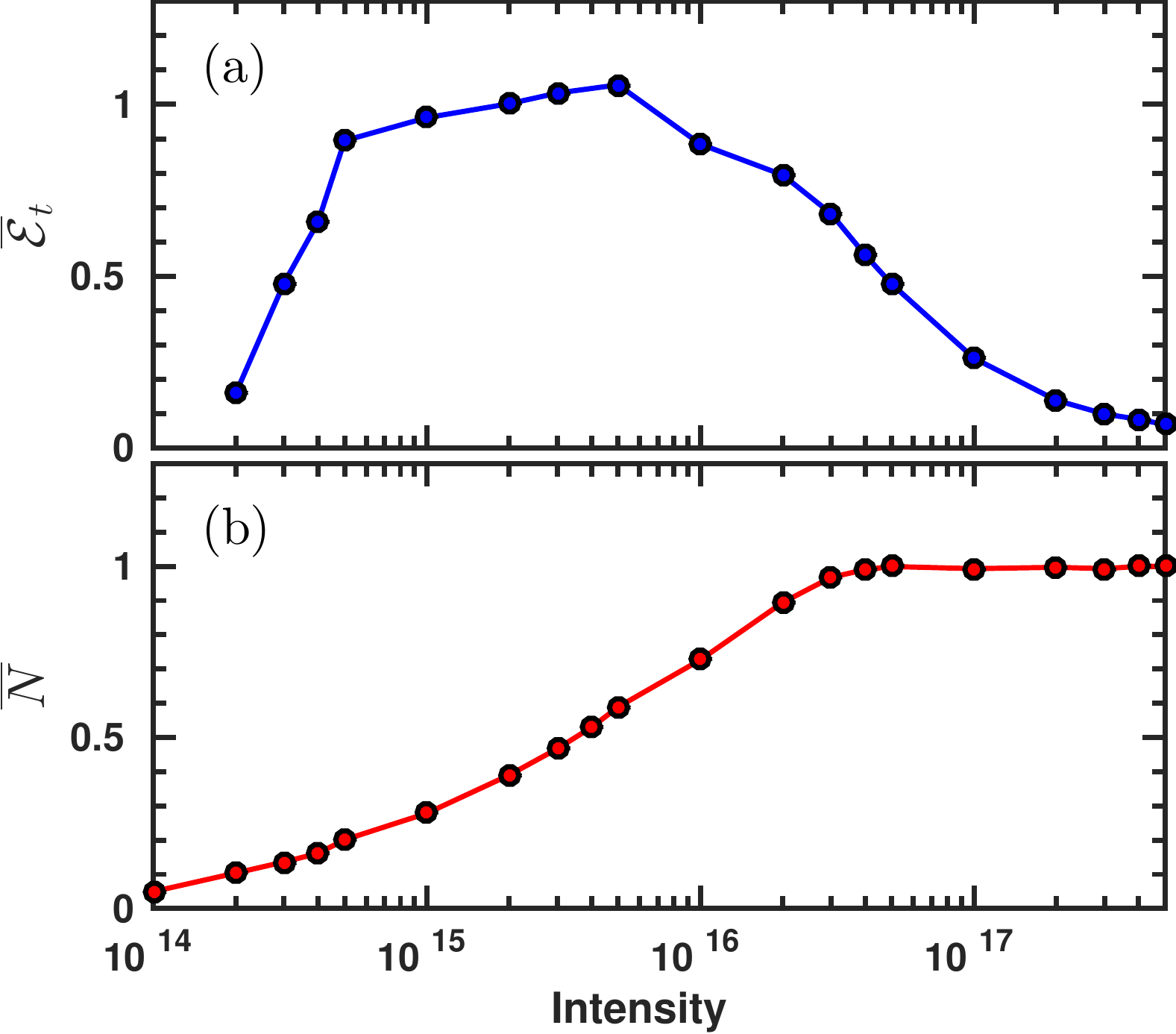}
\caption{(color online) Total absorbed energy per electron in units of $U_p$ (top) and fractional outer ionization (bottom) versus the laser intensity for a deuterium cluster of 
radius $R=2.05$~nm, and number of atoms $N=1791$, density 
$\rho \approx 27.3\rho_c$ and $\omegaM^2/\omega^2 \approx 9.24$. Cluster
is illuminated by a $n=5$-cycle laser pulse of wavelength 
$800$~nm and various peak intensity.} 
\label{fig5}
\end{figure}

\subsection{Analysis of electron trajectory and finding the AHR}
\label{sec3c}
The high level of absorption and outer ionization shown in Fig.\ref{fig5} 
with a short 13.5 fs laser pulse is certainly not due to the 
linear resonance process. Figures~\ref{fig6}(a-b) show normalized space
charge field $\overline{E}_x^{sc} = E_x^{sc}(t)/E_0$ and the
total field $\overline{E}_{x}^t = E_{x}^t(t)/E_0$ versus excursion 
$\overline{x}(t)$ of a few 
selected outer ionized electrons (only 29 electrons are plotted) at the 
peak intensity $5\times 10^{15}\,\Wcmcm$ of Fig.\ref{fig5}. Corresponding $\overline{x}(t)$ versus 
$\overline{t}$ are shown in Fig.\ref{fig6}(c).
The crossing of trajectories of MD electrons [in Fig.\ref{fig6}(c)]
emitted from the cluster at different times, 
their non-laminar motion in time, 
the uncompensated laser field by the space charge field [in Fig.\ref{fig6}(a)] 
and the corresponding non-zero total field [in Fig.\ref{fig6}(b)] 
inside the cluster ($-1\le \overline{x}\le1$) clearly suggest that absorption is not due the 
celebrated Brunel effect \cite{Brunel}.
\begin{figure}[]
\includegraphics[width = 1.0\linewidth,height=0.5\textwidth]{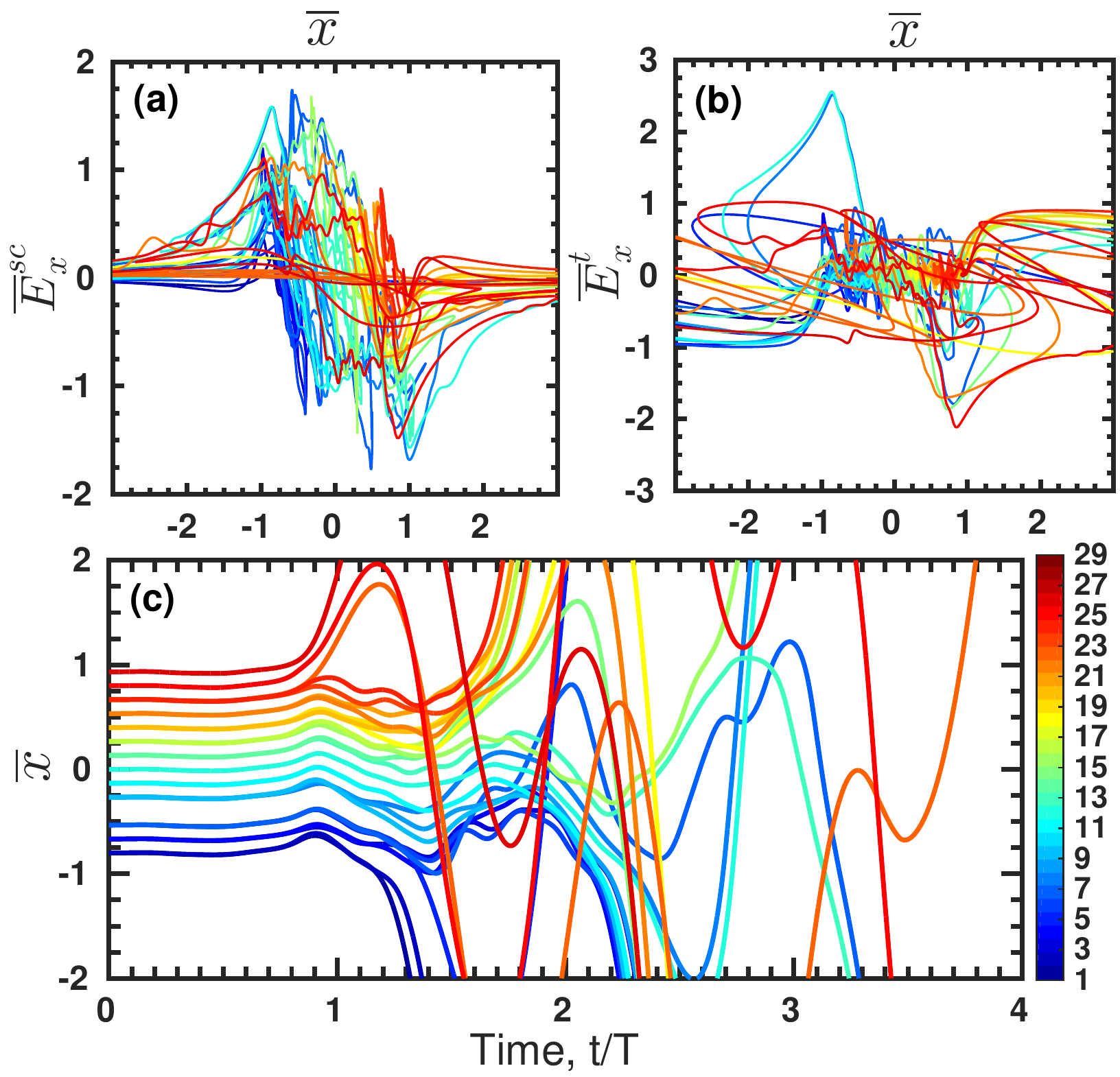}
\caption{(color online) (a) Normalized self-generated space-charge 
field ($E^{sc}(t)/E_0$) and (b) normalized 
total field ($E_{t}(t)/E_0$) versus normalized excursion $x(t)/R$ of a few 
selected outer ionized electrons (only 29 electrons out of $N=1791$ 
are plotted, color bar indicates their index $i=1-29$) at a peak intensity of $5\times 10^{15}\,\Wcmcm$. 
Corresponding trajectories $x(t)/R$ versus $t/T$ are shown in (c).
All other the parameters of cluster and the laser are same as 
in Fig~\ref{fig5}.}
\label{fig6}
\end{figure}  
%
The underlying mechanism can be understood by analyzing trajectories of those MD electrons and finding the corresponding effective frequency as shown by the model in Sec.\ref{sec2b}. 
We write the time dependent frequency of the $i$-th MD electron 
(in analogy with Eq.\reff{EffFreqSq}) as 
\begin{equation}
\Omega^2[r_i(t)] = \frac{\vec{E_{i}^{sc}} (\vekt{r}_i,t) \cdot \vekt{r}_i}{\vekt{r}_i^2} = \frac{\text{restoring field}}{\text{excursion}}
\end{equation} 
where ${\vec{E_i}}^{sc} = \vec{F_i}/q_i$ is the electrostatic field on the $i$-th MD electron obtained from Eq.~\reff{MD_force}.


\begin{figure}[t]  
\includegraphics[width=0.95\linewidth]{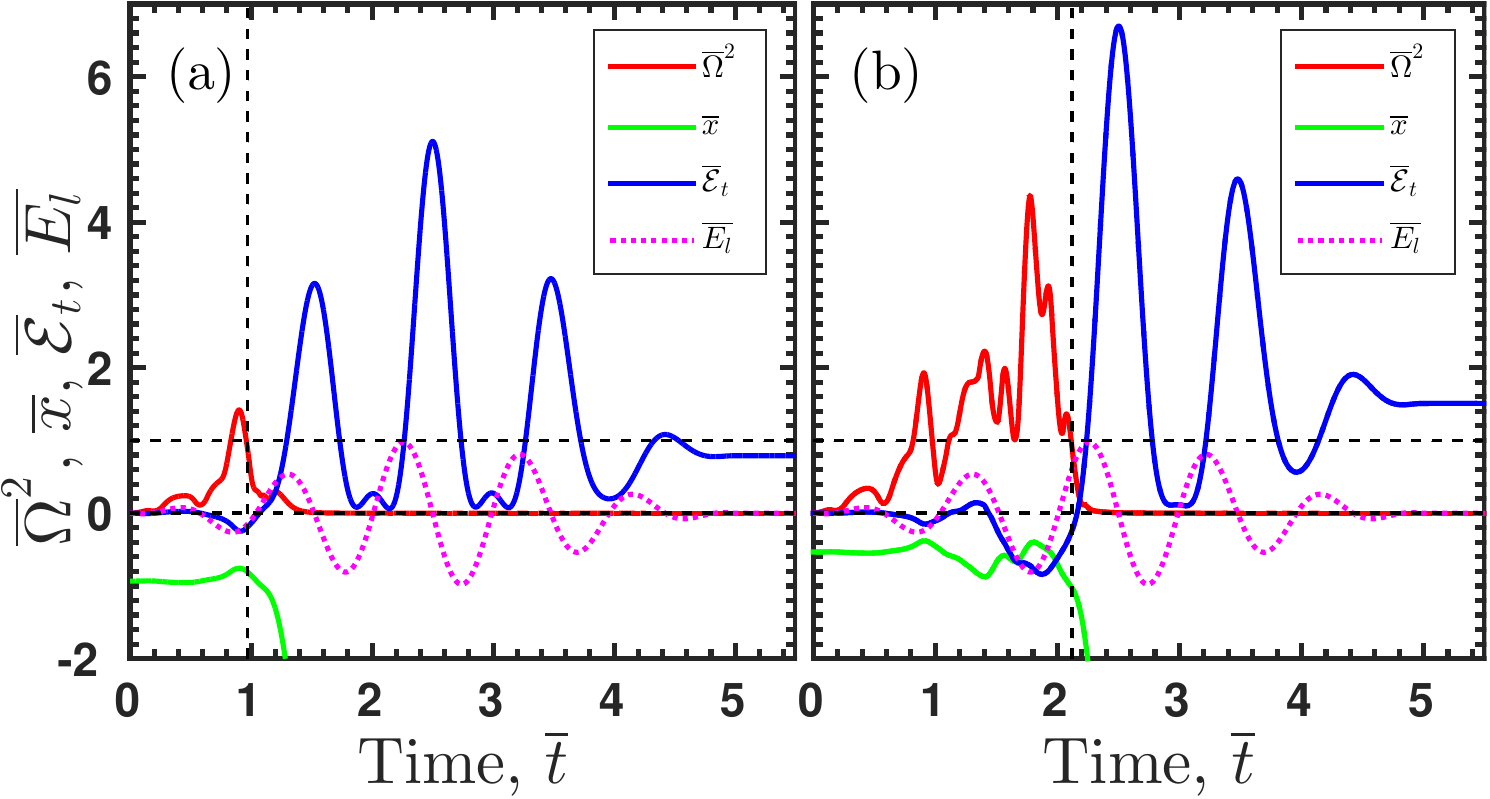}
\caption{(color online) Normalized value of the square of the 
effective frequency $\overline{\Omega}^2(r)$, 
excursion $\overline{x}$, 
total energy $\overline{\mathcal{E}}_t$ 
and the laser field $\overline{E}_l$ versus the normalized 
time $\overline{t}$ 
for different MD electrons undergoing AHR and outer ionization 
at times
$(a)\, \overline{t} = 1.1, (b)\, \overline{t} = 2.2$.  
The cluster is irradiated by a $n = 5$-cycle pulse 
of peak intensity $5 \times 10^{15}\,\mathrm{W/cm^2}$. 
These results resemble with the results of model analysis 
in Fig.~\ref{fig2}.  
\label{fig7}
 }
\end{figure}  

Figures~\ref{fig7}(a-b) show different
normalized quantities, i.e., frequency squared 
$\overline{\Omega}^2$, excursion $\overline{x}$, total 
energy $\overline{\mathcal{E}_t}$, laser field $\overline{E}_l$ versus 
normalized time $\overline{t}$ for selected MD electrons which are 
outer ionized at times: (a) $\overline{t} = 1.1$ 
and (b) $\overline{t} = 2.2$ from the cluster 
irradiated by the same $5$-cycle laser pulse of peak intensity 
$5\times10^{15}\,\Wcmcm$. 
Initially, the cluster is charge neutral, electrostatic field 
is zero inside the cluster and all particles are at rest. As a result, 
the effective frequency $\Omega([r(t)]$ and total energy 
$\overline{\mathcal{E}_t}$ of each electron is zero. As the laser field is 
switched on, the charge separation potential and the corresponding field 
are dynamically created due to the movement of more mobile electrons 
than the slow moving ions. The MD electron in 
Fig.\ref{fig7}(a) is first attracted inside such potential by the restoring force 
due to ions (see its excursion $|\overline{x}|$ 
decreases towards the center of the cluster and total energy starts becoming negative) where 
its effective frequency $\Omega[r(t)]$ after increasing from zero 
exceeds the laser frequency $\omega$ and goes to a maximum value 
when its total energy reaches a minimum negative value 
(i.e, it becomes more bound in the potential).
From this point onwards the dynamics of the MD electron is very similar 
to the electron sphere in the model. 
As the laser field changes further, electron is pulled towards the 
negative $x$-direction, $|\overline{x}|$ increases beyond unity, 
$\overline{\Omega}^2$ drops from its maximum and 
crosses the line of AHR (horizontal dashed line where 
$\overline{\Omega}^2 = 1$) near $t/T \approx 0.95$ with the corresponding 
increase in $\overline{\mathcal{E}}_t$ from negative to positive value 
(bound to free motion) similar to that shown in Fig.\ref{fig2} using 
the model. After the AHR, electron leaves the cluster forever 
with a total energy of $0.8\,U_p$ in the end of the pulse. 
In this early time of interaction, the laser field being very weak, 
only the loosely bound outer most electrons as compared to the core 
electrons leave the cluster. Such early leaving electrons which experience 
AHR in a shallower potential with a low laser field strength 
generally carry low kinetic energies. 


As the laser field increases to it's peak value, more 
electrons are outer ionized from the core of the cluster [as 
in Fig.\ref{fig7}(b)]. 
They experience a relatively deeper potential. Electrons while moving 
in deep potentials have relatively higher $\Omega[r(t)]$ 
[clear from \ref{fig7}(b)] and they require higher field strengths for their 
liberation. Indeed, MD electron in Fig.\ref{fig7}(b) becomes free 
when its $\overline{\Omega}^2$ passes the 
resonance line $\overline{\Omega}^2 = 1$ after dropping from its 
maximum value 
and its energy $\overline{\mathcal{E}}_t$ becomes positive 
at $\overline{t} \approx 2.2$ as the 
peak of the pulse is approached.



The occurrence of AHR for MD electrons in Fig.\ref{fig7} resemble 
with Fig.\ref{fig2} in 
the model in Sec.\ref{sec2} except that frequency and potential start
from zero and self-consistently generated in the case of MD while those 
are predefined in the model.

\begin{figure}[]
\includegraphics[width=1.0\linewidth]{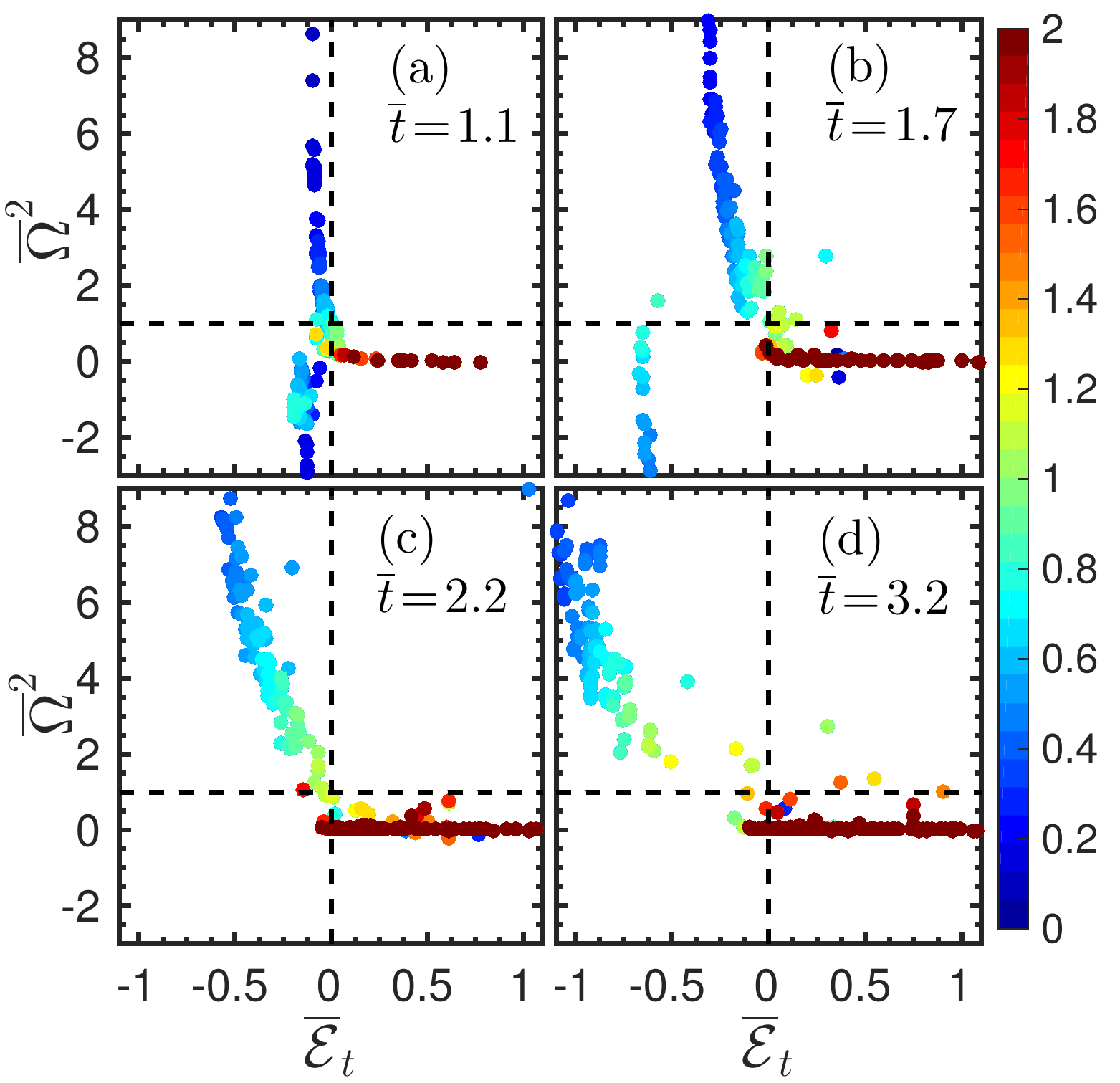}
\caption{(color online) 
Snapshots all MD electrons 
in the (${{\overline{\mathcal{E}}}_{t}}$, $\overline{\Omega}^2$) plane
at times (a) $\overline{t} = 1.1$, (b) $\overline{t} = 1.7$, (c) $\overline{t} = 2.2$, (d) $\overline{t} = 3.2$. 
As the laser field strength increases with time, more and more electrons are drawn towards the line of AHR, i.e, dashed line at $\overline{\Omega}^2 \approx 1$.
The parameters of laser and cluster are same as in Fig.\ref{fig7}. 
 }
\label{fig8}
\end{figure}  

\subsubsection{AHR in the frequency vs energy plane}
\label{NLR_frequencyvsenergy}
To prove that all MD electrons essentially pass through AHR 
during their outer 
ionization, Figures~\ref{fig8}(a-d) show snapshots of all electrons 
in the ($\overline{\Omega}^2$, $\overline{\mathcal{E}}_t$) plane
at different times $\overline{t} = 1.1, 1.7, 2.2, 3.2$ respectively.
Colors indicate normalized positions ($r$) of those electrons 
as in the Fig.\ref{fig3}. From Fig.~\ref{fig8} it is clear 
that each electron leaves the cluster ($r>1$, green to dark red) 
and its energy becomes positive {\em only} when it crosses the
line of AHR (dashed horizontal line at $\overline{\Omega}^2  = 1$). 
After becoming free, electrons have zero effective frequency as they 
are beyond the influence of the electrostatic field. 

In the early time $\overline{t}=1.1$, 
in Fig.\ref{fig8}(a), only few electrons are outer ionized form the cluster and the 
resulting potential is shallow.
As a result energies ($\overline{\mathcal{E}}_t$) of the bound electrons 
are very close to zero but negative. Some of the bound electrons have negative 
${\overline{\Omega}}^2$ due to the repulsion of the compressed electron 
cloud in their vicinity at this early time. 

At later times $\overline{t}=1.7,2.2$, in Figs.\ref{fig8}(b)-(c), 
as the laser field approaches 
its peak value, an increasing number of electrons are outer 
ionized via the AHR channel. 
As a result the potential depth gradually increases, 
remaining bound electrons move to a deeper potential
due to the gradually stronger attractive force of the 
uncompensated bare ionic background, 
the population of negative ${\overline{\Omega}}^2$ 
valued electrons moves gradually to the attractive potential 
(repulsion vanishes with increasing potential depth) 
and becomes almost negligible in Fig.\ref{fig8}(c) 
where all bound electrons are aligned 
to pass the AHR in the next time interval. 

After the peak of the laser pulse, e.g., at $\overline{t}=3.2$ 
in Fig.\ref{fig8}(d),
since outer ionization is mostly saturated and the potential has 
already reached to its near maximum depth at 
the pulse peak before (i.e., near $\overline{t}=2.5$), many bound 
electrons are dragged into the potential 
(they have more negative energies) due to the weakening of the laser field 
compared to the attractive force due to ions.
Some of the quasi-free electrons 
(electrons with positive ${\overline{\Omega}}^2$ and positive 
$\overline{\mathcal{E}}_t$)
near the cluster boundary also 
return inside due to such attraction. 

The feature of AHR shown in Fig.\ref{fig8} in the frequency versus energy 
plane resembles Fig.\ref{fig3} obtained by the model, 
except that some MD electrons in Fig.\ref{fig8} 
experience negative frequencies due to repulsion of the neighbouring
electrons.

Above analysis of trajectories of MD electrons in the self-generated, 
time-varying potential clearly indicates that the passage of AHR is must 
during their outer ionization. The fact that large amount of energy 
absorption by an electron and its simultaneous liberation from the 
dynamical potential happening at the same time {\em only} when AHR 
condition is met, clearly proves the AHR as a responsible mechanism 
behind this efficient laser absorption in a cluster.
 
\section{Summary and conclusion}\label{sec4}
The goal of this work is to re-examine the AHR absorption mechanism of 
intense infrared laser pulses in a over-dense cluster using MD 
simulations. 
Although, AHR was proved earlier by a rigid sphere model and 
particle-in-cell (PIC) simulations \cite{MKunduprl} of clusters, 
it still remains obscure in many-body plasma simulations.
To prove AHR on a firm footing
a three dimensional MD simulation code with soft-core 
Coulomb interactions among charge particles
has been developed. 
By following the trajectory of each 
individual MD electron and identifying its time-dependent frequency 
$\Omega[r(t)]$ in the self-consistent anharmonic 
potential (as in Ref.\cite{MKunduprl}) it is found that electron 
leaves the potential and becomes free only when AHR condition 
$\Omega[r(t)]=\omega$ is met. 
Thus, for the first time, our MD simulation clearly identifies AHR process in the laser cluster interaction. 
A simple anharmonic oscillator model 
is introduced to understand MD results better. 
The model brings out most of the features of MD electrons while 
passing the AHR. Thus, we not only bridge the gap between 
PIC simulations, analytical models and MD calculations but 
also unequivocally prove 
that AHR processes is a universal dominant collisionless mechanism of 
absorption in the short pulse regime or in the early time of longer 
pulses in clusters. 

We believe that AHR mechanism works irrespective of 
the target size at least in the first few nano-layer of the sharp 
over-dense plasma (zero density scale-length) where laser interacts first
and the analysis of electron trajectories presented here may be useful 
to identify AHR in such targets. 

The prompt generation of electrons via AHR within a time much shorter than a laser period, the crossing of electron trajectories 
[in Fig.\ref{fig6}(c)] demonstrated by MD simulations and  
the breaking of laminar flow of electrons may lead to 
plasma wave-breaking and subsequent mixing of wave-phases 
\cite{Mulser2015} even at sub-relativistic laser intensities
in an extended plasma.

\label{conclusion}

\begin{acknowledgments}
Authors would like to thank Sudip Sengupta for careful reading of the manuscript and providing useful suggestions.
\end{acknowledgments}

\nocite{*}
\bibliography{clusterSMMKv1}

\providecommand{\noopsort}[1]{}\providecommand{\singleletter}[1]{#1}%
\begin{thebibliography}{64}%
\makeatletter
\providecommand \@ifxundefined [1]{%
 \@ifx{#1\undefined}
}%
\providecommand \@ifnum [1]{%
 \ifnum #1\expandafter \@firstoftwo
 \else \expandafter \@secondoftwo
 \fi
}%
\providecommand \@ifx [1]{%
 \ifx #1\expandafter \@firstoftwo
 \else \expandafter \@secondoftwo
 \fi
}%
\providecommand \natexlab [1]{#1}%
\providecommand \enquote  [1]{``#1''}%
\providecommand \bibnamefont  [1]{#1}%
\providecommand \bibfnamefont [1]{#1}%
\providecommand \citenamefont [1]{#1}%
\providecommand \href@noop [0]{\@secondoftwo}%
\providecommand \href [0]{\begingroup \@sanitize@url \@href}%
\providecommand \@href[1]{\@@startlink{#1}\@@href}%
\providecommand \@@href[1]{\endgroup#1\@@endlink}%
\providecommand \@sanitize@url [0]{\catcode `\\12\catcode `\$12\catcode
  `\&12\catcode `\#12\catcode `\^12\catcode `\_12\catcode `\%12\relax}%
\providecommand \@@startlink[1]{}%
\providecommand \@@endlink[0]{}%
\providecommand \url  [0]{\begingroup\@sanitize@url \@url }%
\providecommand \@url [1]{\endgroup\@href {#1}{\urlprefix }}%
\providecommand \urlprefix  [0]{URL }%
\providecommand \Eprint [0]{\href }%
\providecommand \doibase [0]{http://dx.doi.org/}%
\providecommand \selectlanguage [0]{\@gobble}%
\providecommand \bibinfo  [0]{\@secondoftwo}%
\providecommand \bibfield  [0]{\@secondoftwo}%
\providecommand \translation [1]{[#1]}%
\providecommand \BibitemOpen [0]{}%
\providecommand \bibitemStop [0]{}%
\providecommand \bibitemNoStop [0]{.\EOS\space}%
\providecommand \EOS [0]{\spacefactor3000\relax}%
\providecommand \BibitemShut  [1]{\csname bibitem#1\endcsname}%
\let\auto@bib@innerbib\@empty
\bibitem [{\citenamefont {Ditmire}\ \emph
  {et~al.}(1997{\natexlab{a}})\citenamefont {Ditmire}, \citenamefont {Smith},
  \citenamefont {Tisch},\ and\ \citenamefont {Hutchinson}}]{Ditmire_PRL78}%
  \BibitemOpen
  \bibfield  {author} {\bibinfo {author} {\bibfnamefont {T.}~\bibnamefont
  {Ditmire}}, \bibinfo {author} {\bibfnamefont {R.~A.}\ \bibnamefont {Smith}},
  \bibinfo {author} {\bibfnamefont {J.~W.~G.}\ \bibnamefont {Tisch}}, \ and\
  \bibinfo {author} {\bibfnamefont {M.~H.~R.}\ \bibnamefont {Hutchinson}},\
  }\href {\doibase 10.1103/PhysRevLett.78.3121} {\bibfield  {journal} {\bibinfo
   {journal} {Phys. Rev. Lett.}\ }\textbf {\bibinfo {volume} {78}},\ \bibinfo
  {pages} {3121} (\bibinfo {year} {1997}{\natexlab{a}})}\BibitemShut {NoStop}%
\bibitem [{\citenamefont {Rose-Petruck}\ \emph {et~al.}(1997)\citenamefont
  {Rose-Petruck}, \citenamefont {Schafer}, \citenamefont {Wilson},\ and\
  \citenamefont {Barty}}]{RosePetruck}%
  \BibitemOpen
  \bibfield  {author} {\bibinfo {author} {\bibfnamefont {C.}~\bibnamefont
  {Rose-Petruck}}, \bibinfo {author} {\bibfnamefont {K.~J.}\ \bibnamefont
  {Schafer}}, \bibinfo {author} {\bibfnamefont {K.~R.}\ \bibnamefont {Wilson}},
  \ and\ \bibinfo {author} {\bibfnamefont {C.~P.~J.}\ \bibnamefont {Barty}},\
  }\href {\doibase 10.1103/PhysRevA.55.1182} {\bibfield  {journal} {\bibinfo
  {journal} {Phys. Rev. A}\ }\textbf {\bibinfo {volume} {55}},\ \bibinfo
  {pages} {1182} (\bibinfo {year} {1997})}\BibitemShut {NoStop}%
\bibitem [{\citenamefont {Ditmire}\ \emph
  {et~al.}(1997{\natexlab{b}})\citenamefont {Ditmire}, \citenamefont {Tish},
  \citenamefont {Springate}, \citenamefont {Mason}, \citenamefont {Hay},
  \citenamefont {Marangos},\ and\ \citenamefont
  {Hutchinson}}]{Ditmire_Nature386}%
  \BibitemOpen
  \bibfield  {author} {\bibinfo {author} {\bibfnamefont {T.}~\bibnamefont
  {Ditmire}}, \bibinfo {author} {\bibfnamefont {J.~W.~G.}\ \bibnamefont
  {Tish}}, \bibinfo {author} {\bibfnamefont {E.}~\bibnamefont {Springate}},
  \bibinfo {author} {\bibfnamefont {M.~B.}\ \bibnamefont {Mason}}, \bibinfo
  {author} {\bibfnamefont {N.}~\bibnamefont {Hay}}, \bibinfo {author}
  {\bibfnamefont {J.}~\bibnamefont {Marangos}}, \ and\ \bibinfo {author}
  {\bibfnamefont {M.~H.~R.}\ \bibnamefont {Hutchinson}},\ }\href {\doibase
  10.1038/386054a0} {\bibfield  {journal} {\bibinfo  {journal} {Nature
  (London)}\ }\textbf {\bibinfo {volume} {386}},\ \bibinfo {pages} {54}
  (\bibinfo {year} {1997}{\natexlab{b}})}\BibitemShut {NoStop}%
\bibitem [{\citenamefont {Ditmire}\ \emph
  {et~al.}(1997{\natexlab{c}})\citenamefont {Ditmire}, \citenamefont {Tisch},
  \citenamefont {Springate}, \citenamefont {Mason}, \citenamefont {Hay},
  \citenamefont {Marangos},\ and\ \citenamefont
  {Hutchinson}}]{Ditmire_PRL78_2732}%
  \BibitemOpen
  \bibfield  {author} {\bibinfo {author} {\bibfnamefont {T.}~\bibnamefont
  {Ditmire}}, \bibinfo {author} {\bibfnamefont {J.~W.~G.}\ \bibnamefont
  {Tisch}}, \bibinfo {author} {\bibfnamefont {E.}~\bibnamefont {Springate}},
  \bibinfo {author} {\bibfnamefont {M.~B.}\ \bibnamefont {Mason}}, \bibinfo
  {author} {\bibfnamefont {N.}~\bibnamefont {Hay}}, \bibinfo {author}
  {\bibfnamefont {J.~P.}\ \bibnamefont {Marangos}}, \ and\ \bibinfo {author}
  {\bibfnamefont {M.~H.~R.}\ \bibnamefont {Hutchinson}},\ }\href {\doibase
  10.1103/PhysRevLett.78.2732} {\bibfield  {journal} {\bibinfo  {journal}
  {Phys. Rev. Lett.}\ }\textbf {\bibinfo {volume} {78}},\ \bibinfo {pages}
  {2732} (\bibinfo {year} {1997}{\natexlab{c}})}\BibitemShut {NoStop}%
\bibitem [{\citenamefont {Kumarappan}\ \emph
  {et~al.}(2001{\natexlab{a}})\citenamefont {Kumarappan}, \citenamefont
  {Krishnamurthy},\ and\ \citenamefont {Mathur}}]{Kumarappan_PRL87}%
  \BibitemOpen
  \bibfield  {author} {\bibinfo {author} {\bibfnamefont {V.}~\bibnamefont
  {Kumarappan}}, \bibinfo {author} {\bibfnamefont {M.}~\bibnamefont
  {Krishnamurthy}}, \ and\ \bibinfo {author} {\bibfnamefont {D.}~\bibnamefont
  {Mathur}},\ }\href {\doibase 10.1103/PhysRevLett.87.085005} {\bibfield
  {journal} {\bibinfo  {journal} {Phys. Rev. Lett.}\ }\textbf {\bibinfo
  {volume} {87}},\ \bibinfo {pages} {085005} (\bibinfo {year}
  {2001}{\natexlab{a}})}\BibitemShut {NoStop}%
\bibitem [{\citenamefont {Lezius}\ \emph {et~al.}(1998)\citenamefont {Lezius},
  \citenamefont {Dobosz}, \citenamefont {Normand},\ and\ \citenamefont
  {Schmidt}}]{Lezius}%
  \BibitemOpen
  \bibfield  {author} {\bibinfo {author} {\bibfnamefont {M.}~\bibnamefont
  {Lezius}}, \bibinfo {author} {\bibfnamefont {S.}~\bibnamefont {Dobosz}},
  \bibinfo {author} {\bibfnamefont {D.}~\bibnamefont {Normand}}, \ and\
  \bibinfo {author} {\bibfnamefont {M.}~\bibnamefont {Schmidt}},\ }\href
  {\doibase 10.1103/PhysRevLett.80.261} {\bibfield  {journal} {\bibinfo
  {journal} {Phys. Rev. Lett.}\ }\textbf {\bibinfo {volume} {80}},\ \bibinfo
  {pages} {261} (\bibinfo {year} {1998})}\BibitemShut {NoStop}%
\bibitem [{\citenamefont {Fukuda}\ \emph {et~al.}(2003)\citenamefont {Fukuda},
  \citenamefont {Yamakawa}, \citenamefont {Akahane}, \citenamefont {Aoyama},
  \citenamefont {Inoue}, \citenamefont {Ueda},\ and\ \citenamefont
  {Kishimoto}}]{Fukuda}%
  \BibitemOpen
  \bibfield  {author} {\bibinfo {author} {\bibfnamefont {Y.}~\bibnamefont
  {Fukuda}}, \bibinfo {author} {\bibfnamefont {K.}~\bibnamefont {Yamakawa}},
  \bibinfo {author} {\bibfnamefont {Y.}~\bibnamefont {Akahane}}, \bibinfo
  {author} {\bibfnamefont {M.}~\bibnamefont {Aoyama}}, \bibinfo {author}
  {\bibfnamefont {N.}~\bibnamefont {Inoue}}, \bibinfo {author} {\bibfnamefont
  {H.}~\bibnamefont {Ueda}}, \ and\ \bibinfo {author} {\bibfnamefont
  {Y.}~\bibnamefont {Kishimoto}},\ }\href {\doibase 10.1103/PhysRevA.67.061201}
  {\bibfield  {journal} {\bibinfo  {journal} {Phys. Rev. A}\ }\textbf {\bibinfo
  {volume} {67}},\ \bibinfo {pages} {061201} (\bibinfo {year}
  {2003})}\BibitemShut {NoStop}%
\bibitem [{\citenamefont {Kumarappan}\ \emph
  {et~al.}(2001{\natexlab{b}})\citenamefont {Kumarappan}, \citenamefont
  {Krishnamurthy}, \citenamefont {Mathur},\ and\ \citenamefont
  {Tribedi}}]{Kumarappan2001}%
  \BibitemOpen
  \bibfield  {author} {\bibinfo {author} {\bibfnamefont {V.}~\bibnamefont
  {Kumarappan}}, \bibinfo {author} {\bibfnamefont {M.}~\bibnamefont
  {Krishnamurthy}}, \bibinfo {author} {\bibfnamefont {D.}~\bibnamefont
  {Mathur}}, \ and\ \bibinfo {author} {\bibfnamefont {L.~C.}\ \bibnamefont
  {Tribedi}},\ }\href {\doibase 10.1103/PhysRevA.63.023203} {\bibfield
  {journal} {\bibinfo  {journal} {Phys. Rev. A}\ }\textbf {\bibinfo {volume}
  {63}},\ \bibinfo {pages} {023203} (\bibinfo {year}
  {2001}{\natexlab{b}})}\BibitemShut {NoStop}%
\bibitem [{\citenamefont {Krishnamurthy}\ \emph {et~al.}(2004)\citenamefont
  {Krishnamurthy}, \citenamefont {Mathur},\ and\ \citenamefont
  {Kumarappan}}]{Krishnamurthy}%
  \BibitemOpen
  \bibfield  {author} {\bibinfo {author} {\bibfnamefont {M.}~\bibnamefont
  {Krishnamurthy}}, \bibinfo {author} {\bibfnamefont {D.}~\bibnamefont
  {Mathur}}, \ and\ \bibinfo {author} {\bibfnamefont {V.}~\bibnamefont
  {Kumarappan}},\ }\href {\doibase 10.1103/PhysRevA.69.033202} {\bibfield
  {journal} {\bibinfo  {journal} {Phys. Rev. A}\ }\textbf {\bibinfo {volume}
  {69}},\ \bibinfo {pages} {033202} (\bibinfo {year} {2004})}\BibitemShut
  {NoStop}%
\bibitem [{\citenamefont {Kumarappan}\ \emph {et~al.}(2002)\citenamefont
  {Kumarappan}, \citenamefont {Krishnamurthy},\ and\ \citenamefont
  {Mathur}}]{Kumarappan2002}%
  \BibitemOpen
  \bibfield  {author} {\bibinfo {author} {\bibfnamefont {V.}~\bibnamefont
  {Kumarappan}}, \bibinfo {author} {\bibfnamefont {M.}~\bibnamefont
  {Krishnamurthy}}, \ and\ \bibinfo {author} {\bibfnamefont {D.}~\bibnamefont
  {Mathur}},\ }\href {\doibase 10.1103/PhysRevA.66.033203} {\bibfield
  {journal} {\bibinfo  {journal} {Phys. Rev. A}\ }\textbf {\bibinfo {volume}
  {66}},\ \bibinfo {pages} {033203} (\bibinfo {year} {2002})}\BibitemShut
  {NoStop}%
\bibitem [{\citenamefont {Ditmire}\ \emph {et~al.}(1998)\citenamefont
  {Ditmire}, \citenamefont {Springate}, \citenamefont {Tisch}, \citenamefont
  {Shao}, \citenamefont {Mason}, \citenamefont {Hay}, \citenamefont
  {Marangos},\ and\ \citenamefont {Hutchinson}}]{Ditmire_PRA57}%
  \BibitemOpen
  \bibfield  {author} {\bibinfo {author} {\bibfnamefont {T.}~\bibnamefont
  {Ditmire}}, \bibinfo {author} {\bibfnamefont {E.}~\bibnamefont {Springate}},
  \bibinfo {author} {\bibfnamefont {J.~W.~G.}\ \bibnamefont {Tisch}}, \bibinfo
  {author} {\bibfnamefont {Y.~L.}\ \bibnamefont {Shao}}, \bibinfo {author}
  {\bibfnamefont {M.~B.}\ \bibnamefont {Mason}}, \bibinfo {author}
  {\bibfnamefont {N.}~\bibnamefont {Hay}}, \bibinfo {author} {\bibfnamefont
  {J.~P.}\ \bibnamefont {Marangos}}, \ and\ \bibinfo {author} {\bibfnamefont
  {M.~H.~R.}\ \bibnamefont {Hutchinson}},\ }\href {\doibase
  10.1103/PhysRevA.57.369} {\bibfield  {journal} {\bibinfo  {journal} {Phys.
  Rev. A}\ }\textbf {\bibinfo {volume} {57}},\ \bibinfo {pages} {369} (\bibinfo
  {year} {1998})}\BibitemShut {NoStop}%
\bibitem [{\citenamefont {Springate}\ \emph {et~al.}(2000)\citenamefont
  {Springate}, \citenamefont {Hay}, \citenamefont {Tisch}, \citenamefont
  {Mason}, \citenamefont {Ditmire}, \citenamefont {Hutchinson},\ and\
  \citenamefont {Marangos}}]{Springate_PRA61}%
  \BibitemOpen
  \bibfield  {author} {\bibinfo {author} {\bibfnamefont {E.}~\bibnamefont
  {Springate}}, \bibinfo {author} {\bibfnamefont {N.}~\bibnamefont {Hay}},
  \bibinfo {author} {\bibfnamefont {J.~W.~G.}\ \bibnamefont {Tisch}}, \bibinfo
  {author} {\bibfnamefont {M.~B.}\ \bibnamefont {Mason}}, \bibinfo {author}
  {\bibfnamefont {T.}~\bibnamefont {Ditmire}}, \bibinfo {author} {\bibfnamefont
  {M.~H.~R.}\ \bibnamefont {Hutchinson}}, \ and\ \bibinfo {author}
  {\bibfnamefont {J.~P.}\ \bibnamefont {Marangos}},\ }\href {\doibase
  10.1103/PhysRevA.61.063201} {\bibfield  {journal} {\bibinfo  {journal} {Phys.
  Rev. A}\ }\textbf {\bibinfo {volume} {61}},\ \bibinfo {pages} {063201}
  (\bibinfo {year} {2000})}\BibitemShut {NoStop}%
\bibitem [{\citenamefont {Springate}\ \emph {et~al.}(2003)\citenamefont
  {Springate}, \citenamefont {Aseyev}, \citenamefont {Zamith},\ and\
  \citenamefont {Vrakking}}]{Springate_PRA68}%
  \BibitemOpen
  \bibfield  {author} {\bibinfo {author} {\bibfnamefont {E.}~\bibnamefont
  {Springate}}, \bibinfo {author} {\bibfnamefont {S.~A.}\ \bibnamefont
  {Aseyev}}, \bibinfo {author} {\bibfnamefont {S.}~\bibnamefont {Zamith}}, \
  and\ \bibinfo {author} {\bibfnamefont {M.~J.~J.}\ \bibnamefont {Vrakking}},\
  }\href {\doibase 10.1103/PhysRevA.68.053201} {\bibfield  {journal} {\bibinfo
  {journal} {Phys. Rev. A}\ }\textbf {\bibinfo {volume} {68}},\ \bibinfo
  {pages} {053201} (\bibinfo {year} {2003})}\BibitemShut {NoStop}%
\bibitem [{\citenamefont {Shao}\ \emph {et~al.}(1996)\citenamefont {Shao},
  \citenamefont {Ditmire}, \citenamefont {Tisch}, \citenamefont {Springate},
  \citenamefont {Marangos},\ and\ \citenamefont {Hutchinson}}]{Shao_PRL77}%
  \BibitemOpen
  \bibfield  {author} {\bibinfo {author} {\bibfnamefont {Y.~L.}\ \bibnamefont
  {Shao}}, \bibinfo {author} {\bibfnamefont {T.}~\bibnamefont {Ditmire}},
  \bibinfo {author} {\bibfnamefont {J.~W.~G.}\ \bibnamefont {Tisch}}, \bibinfo
  {author} {\bibfnamefont {E.}~\bibnamefont {Springate}}, \bibinfo {author}
  {\bibfnamefont {J.~P.}\ \bibnamefont {Marangos}}, \ and\ \bibinfo {author}
  {\bibfnamefont {M.~H.~R.}\ \bibnamefont {Hutchinson}},\ }\href {\doibase
  10.1103/PhysRevLett.77.3343} {\bibfield  {journal} {\bibinfo  {journal}
  {Phys. Rev. Lett.}\ }\textbf {\bibinfo {volume} {77}},\ \bibinfo {pages}
  {3343} (\bibinfo {year} {1996})}\BibitemShut {NoStop}%
\bibitem [{\citenamefont {Chen}\ \emph {et~al.}(2002)\citenamefont {Chen},
  \citenamefont {Park}, \citenamefont {Hong}, \citenamefont {Choi},
  \citenamefont {Kim}, \citenamefont {Zhang},\ and\ \citenamefont
  {Nam}}]{Chen_POP_9}%
  \BibitemOpen
  \bibfield  {author} {\bibinfo {author} {\bibfnamefont {L.~M.}\ \bibnamefont
  {Chen}}, \bibinfo {author} {\bibfnamefont {J.~J.}\ \bibnamefont {Park}},
  \bibinfo {author} {\bibfnamefont {K.~H.}\ \bibnamefont {Hong}}, \bibinfo
  {author} {\bibfnamefont {I.~W.}\ \bibnamefont {Choi}}, \bibinfo {author}
  {\bibfnamefont {J.~L.}\ \bibnamefont {Kim}}, \bibinfo {author} {\bibfnamefont
  {J.}~\bibnamefont {Zhang}}, \ and\ \bibinfo {author} {\bibfnamefont {C.~H.}\
  \bibnamefont {Nam}},\ }\href {\doibase http://dx.doi.org/10.1063/1.1492804}
  {\bibfield  {journal} {\bibinfo  {journal} {Physics of Plasmas}\ }\textbf
  {\bibinfo {volume} {9}},\ \bibinfo {pages} {3595} (\bibinfo {year}
  {2002})}\BibitemShut {NoStop}%
\bibitem [{\citenamefont {Kumarappan}\ \emph {et~al.}(2003)\citenamefont
  {Kumarappan}, \citenamefont {Krishnamurthy},\ and\ \citenamefont
  {Mathur}}]{Kumarappan2003}%
  \BibitemOpen
  \bibfield  {author} {\bibinfo {author} {\bibfnamefont {V.}~\bibnamefont
  {Kumarappan}}, \bibinfo {author} {\bibfnamefont {M.}~\bibnamefont
  {Krishnamurthy}}, \ and\ \bibinfo {author} {\bibfnamefont {D.}~\bibnamefont
  {Mathur}},\ }\href {\doibase 10.1103/PhysRevA.67.043204} {\bibfield
  {journal} {\bibinfo  {journal} {Phys. Rev. A}\ }\textbf {\bibinfo {volume}
  {67}},\ \bibinfo {pages} {043204} (\bibinfo {year} {2003})}\BibitemShut
  {NoStop}%
\bibitem [{\citenamefont {Jha}\ \emph {et~al.}(2005)\citenamefont {Jha},
  \citenamefont {Mathur},\ and\ \citenamefont {Krishnamurthy}}]{Jha_2005}%
  \BibitemOpen
  \bibfield  {author} {\bibinfo {author} {\bibfnamefont {J.}~\bibnamefont
  {Jha}}, \bibinfo {author} {\bibfnamefont {D.}~\bibnamefont {Mathur}}, \ and\
  \bibinfo {author} {\bibfnamefont {M.}~\bibnamefont {Krishnamurthy}},\ }\href
  {http://stacks.iop.org/0953-4075/38/i=18/a=L01} {\bibfield  {journal}
  {\bibinfo  {journal} {Journal of Physics B: Atomic, Molecular and Optical
  Physics}\ }\textbf {\bibinfo {volume} {38}},\ \bibinfo {pages} {L291}
  (\bibinfo {year} {2005})}\BibitemShut {NoStop}%
\bibitem [{\citenamefont {Jha}\ \emph {et~al.}(2006)\citenamefont {Jha},
  \citenamefont {Mathur},\ and\ \citenamefont {Krishnamurthy}}]{Jha_2006}%
  \BibitemOpen
  \bibfield  {author} {\bibinfo {author} {\bibfnamefont {J.}~\bibnamefont
  {Jha}}, \bibinfo {author} {\bibfnamefont {D.}~\bibnamefont {Mathur}}, \ and\
  \bibinfo {author} {\bibfnamefont {M.}~\bibnamefont {Krishnamurthy}},\ }\href
  {http://scitation.aip.org/content/aip/journal/apl/88/4/10.1063/1.2164414}
  {\bibfield  {journal} {\bibinfo  {journal} {Applied Physics Letters}\
  }\textbf {\bibinfo {volume} {88}},\ \bibinfo {eid} {041107} (\bibinfo {year}
  {2006})}\BibitemShut {NoStop}%
\bibitem [{\citenamefont {Chen}\ \emph {et~al.}(2010)\citenamefont {Chen},
  \citenamefont {Liu}, \citenamefont {Wang}, \citenamefont {Kando},
  \citenamefont {Mao}, \citenamefont {Zhang}, \citenamefont {Ma}, \citenamefont
  {Li}, \citenamefont {Bulanov}, \citenamefont {Tajima}, \citenamefont {Kato},
  \citenamefont {Sheng}, \citenamefont {Wei},\ and\ \citenamefont
  {Zhang}}]{Chen_PRL104}%
  \BibitemOpen
  \bibfield  {author} {\bibinfo {author} {\bibfnamefont {L.~M.}\ \bibnamefont
  {Chen}}, \bibinfo {author} {\bibfnamefont {F.}~\bibnamefont {Liu}}, \bibinfo
  {author} {\bibfnamefont {W.~M.}\ \bibnamefont {Wang}}, \bibinfo {author}
  {\bibfnamefont {M.}~\bibnamefont {Kando}}, \bibinfo {author} {\bibfnamefont
  {J.~Y.}\ \bibnamefont {Mao}}, \bibinfo {author} {\bibfnamefont
  {L.}~\bibnamefont {Zhang}}, \bibinfo {author} {\bibfnamefont {J.~L.}\
  \bibnamefont {Ma}}, \bibinfo {author} {\bibfnamefont {Y.~T.}\ \bibnamefont
  {Li}}, \bibinfo {author} {\bibfnamefont {S.~V.}\ \bibnamefont {Bulanov}},
  \bibinfo {author} {\bibfnamefont {T.}~\bibnamefont {Tajima}}, \bibinfo
  {author} {\bibfnamefont {Y.}~\bibnamefont {Kato}}, \bibinfo {author}
  {\bibfnamefont {Z.~M.}\ \bibnamefont {Sheng}}, \bibinfo {author}
  {\bibfnamefont {Z.~Y.}\ \bibnamefont {Wei}}, \ and\ \bibinfo {author}
  {\bibfnamefont {J.}~\bibnamefont {Zhang}},\ }\href {\doibase
  10.1103/PhysRevLett.104.215004} {\bibfield  {journal} {\bibinfo  {journal}
  {Phys. Rev. Lett.}\ }\textbf {\bibinfo {volume} {104}},\ \bibinfo {pages}
  {215004} (\bibinfo {year} {2010})}\BibitemShut {NoStop}%
\bibitem [{\citenamefont {McPherson}\ \emph {et~al.}(1994)\citenamefont
  {McPherson}, \citenamefont {Thompson}, \citenamefont {Borisov}, \citenamefont
  {Boyer}, ,\ and\ \citenamefont {Rhodes}}]{McPherson_Nature370}%
  \BibitemOpen
  \bibfield  {author} {\bibinfo {author} {\bibfnamefont {A.}~\bibnamefont
  {McPherson}}, \bibinfo {author} {\bibfnamefont {B.~D.}\ \bibnamefont
  {Thompson}}, \bibinfo {author} {\bibfnamefont {A.~B.}\ \bibnamefont
  {Borisov}}, \bibinfo {author} {\bibfnamefont {K.}~\bibnamefont {Boyer}}, , \
  and\ \bibinfo {author} {\bibfnamefont {C.~K.}\ \bibnamefont {Rhodes}},\
  }\href {\doibase 10.1038/370631a0} {\bibfield  {journal} {\bibinfo  {journal}
  {Nature}\ }\textbf {\bibinfo {volume} {370}},\ \bibinfo {pages} {631}
  (\bibinfo {year} {1994})}\BibitemShut {NoStop}%
\bibitem [{\citenamefont {Rajeev}\ \emph {et~al.}(2013)\citenamefont {Rajeev},
  \citenamefont {Trivikram}, \citenamefont {Rishad}, \citenamefont {Narayanan},
  \citenamefont {Krishnakumar},\ and\ \citenamefont
  {Krishnamurthy}}]{Rajeev_Nature}%
  \BibitemOpen
  \bibfield  {author} {\bibinfo {author} {\bibfnamefont {R.}~\bibnamefont
  {Rajeev}}, \bibinfo {author} {\bibfnamefont {T.~M.}\ \bibnamefont
  {Trivikram}}, \bibinfo {author} {\bibfnamefont {K.~P.~M.}\ \bibnamefont
  {Rishad}}, \bibinfo {author} {\bibfnamefont {V.}~\bibnamefont {Narayanan}},
  \bibinfo {author} {\bibfnamefont {E.}~\bibnamefont {Krishnakumar}}, \ and\
  \bibinfo {author} {\bibfnamefont {M.}~\bibnamefont {Krishnamurthy}},\ }\href
  {\doibase 10.1038/nphys2526} {\bibfield  {journal} {\bibinfo  {journal} {Nat
  Phys.}\ }\textbf {\bibinfo {volume} {9}},\ \bibinfo {pages} {185} (\bibinfo
  {year} {2013})}\BibitemShut {NoStop}%
\bibitem [{\citenamefont {Zweiback}\ \emph {et~al.}(1999)\citenamefont
  {Zweiback}, \citenamefont {Ditmire},\ and\ \citenamefont {Perry}}]{Zweiback}%
  \BibitemOpen
  \bibfield  {author} {\bibinfo {author} {\bibfnamefont {J.}~\bibnamefont
  {Zweiback}}, \bibinfo {author} {\bibfnamefont {T.}~\bibnamefont {Ditmire}}, \
  and\ \bibinfo {author} {\bibfnamefont {M.~D.}\ \bibnamefont {Perry}},\ }\href
  {\doibase 10.1103/PhysRevA.59.R3166} {\bibfield  {journal} {\bibinfo
  {journal} {Phys. Rev. A}\ }\textbf {\bibinfo {volume} {59}},\ \bibinfo
  {pages} {R3166} (\bibinfo {year} {1999})}\BibitemShut {NoStop}%
\bibitem [{\citenamefont {Saalmann}\ \emph {et~al.}(2006)\citenamefont
  {Saalmann}, \citenamefont {Siedschlag},\ and\ \citenamefont
  {Rost}}]{Saalmann_JPB39}%
  \BibitemOpen
  \bibfield  {author} {\bibinfo {author} {\bibfnamefont {U.}~\bibnamefont
  {Saalmann}}, \bibinfo {author} {\bibfnamefont {C.}~\bibnamefont
  {Siedschlag}}, \ and\ \bibinfo {author} {\bibfnamefont {J.~M.}\ \bibnamefont
  {Rost}},\ }\href {http://stacks.iop.org/0953-4075/39/i=4/a=R01} {\bibfield
  {journal} {\bibinfo  {journal} {Journal of Physics B: Atomic, Molecular and
  Optical Physics}\ }\textbf {\bibinfo {volume} {39}},\ \bibinfo {pages} {R39}
  (\bibinfo {year} {2006})}\BibitemShut {NoStop}%
\bibitem [{\citenamefont {Fennel}\ \emph {et~al.}(2010)\citenamefont {Fennel},
  \citenamefont {Meiwes-Broer}, \citenamefont {Tiggesb\"aumker}, \citenamefont
  {Reinhard}, \citenamefont {Dinh},\ and\ \citenamefont {Suraud}}]{Fennel_RMP}%
  \BibitemOpen
  \bibfield  {author} {\bibinfo {author} {\bibfnamefont {T.}~\bibnamefont
  {Fennel}}, \bibinfo {author} {\bibfnamefont {K.-H.}\ \bibnamefont
  {Meiwes-Broer}}, \bibinfo {author} {\bibfnamefont {J.}~\bibnamefont
  {Tiggesb\"aumker}}, \bibinfo {author} {\bibfnamefont {P.-G.}\ \bibnamefont
  {Reinhard}}, \bibinfo {author} {\bibfnamefont {P.~M.}\ \bibnamefont {Dinh}},
  \ and\ \bibinfo {author} {\bibfnamefont {E.}~\bibnamefont {Suraud}},\ }\href
  {\doibase 10.1103/RevModPhys.82.1793} {\bibfield  {journal} {\bibinfo
  {journal} {Rev. Mod. Phys.}\ }\textbf {\bibinfo {volume} {82}},\ \bibinfo
  {pages} {1793} (\bibinfo {year} {2010})}\BibitemShut {NoStop}%
\bibitem [{\citenamefont {Ditmire}\ \emph {et~al.}(1996)\citenamefont
  {Ditmire}, \citenamefont {Donnelly}, \citenamefont {Rubenchik}, \citenamefont
  {Falcone},\ and\ \citenamefont {Perry}}]{Ditmire_PRA53}%
  \BibitemOpen
  \bibfield  {author} {\bibinfo {author} {\bibfnamefont {T.}~\bibnamefont
  {Ditmire}}, \bibinfo {author} {\bibfnamefont {T.}~\bibnamefont {Donnelly}},
  \bibinfo {author} {\bibfnamefont {A.~M.}\ \bibnamefont {Rubenchik}}, \bibinfo
  {author} {\bibfnamefont {R.~W.}\ \bibnamefont {Falcone}}, \ and\ \bibinfo
  {author} {\bibfnamefont {M.~D.}\ \bibnamefont {Perry}},\ }\href {\doibase
  10.1103/PhysRevA.53.3379} {\bibfield  {journal} {\bibinfo  {journal} {Phys.
  Rev. A}\ }\textbf {\bibinfo {volume} {53}},\ \bibinfo {pages} {3379}
  (\bibinfo {year} {1996})}\BibitemShut {NoStop}%
\bibitem [{\citenamefont {Kruer}\ and\ \citenamefont
  {Estabrook}(1985)}]{Kruer}%
  \BibitemOpen
  \bibfield  {author} {\bibinfo {author} {\bibfnamefont {W.~L.}\ \bibnamefont
  {Kruer}}\ and\ \bibinfo {author} {\bibfnamefont {K.}~\bibnamefont
  {Estabrook}},\ }\href {\doibase http://dx.doi.org/10.1063/1.865171}
  {\bibfield  {journal} {\bibinfo  {journal} {Physics of Fluids}\ }\textbf
  {\bibinfo {volume} {28}},\ \bibinfo {pages} {430} (\bibinfo {year}
  {1985})}\BibitemShut {NoStop}%
\bibitem [{\citenamefont {Brunel}(1987)}]{Brunel}%
  \BibitemOpen
  \bibfield  {author} {\bibinfo {author} {\bibfnamefont {F.}~\bibnamefont
  {Brunel}},\ }\href {\doibase 10.1103/PhysRevLett.59.52} {\bibfield  {journal}
  {\bibinfo  {journal} {Phys. Rev. Lett.}\ }\textbf {\bibinfo {volume} {59}},\
  \bibinfo {pages} {52} (\bibinfo {year} {1987})}\BibitemShut {NoStop}%
\bibitem [{\citenamefont {Mulser}\ \emph {et~al.}(2012)\citenamefont {Mulser},
  \citenamefont {Weng},\ and\ \citenamefont {Liseykina}}]{Mulser2012}%
  \BibitemOpen
  \bibfield  {author} {\bibinfo {author} {\bibfnamefont {P.}~\bibnamefont
  {Mulser}}, \bibinfo {author} {\bibfnamefont {S.~M.}\ \bibnamefont {Weng}}, \
  and\ \bibinfo {author} {\bibfnamefont {T.}~\bibnamefont {Liseykina}},\ }\href
  {http://scitation.aip.org/content/aip/journal/pop/19/4/10.1063/1.3696034}
  {\bibfield  {journal} {\bibinfo  {journal} {Physics of Plasmas}\ }\textbf
  {\bibinfo {volume} {19}},\ \bibinfo {eid} {043301} (\bibinfo {year}
  {2012})}\BibitemShut {NoStop}%
\bibitem [{\citenamefont {Bauer}\ and\ \citenamefont
  {Macchi}(2003)}]{Bauer2003}%
  \BibitemOpen
  \bibfield  {author} {\bibinfo {author} {\bibfnamefont {D.}~\bibnamefont
  {Bauer}}\ and\ \bibinfo {author} {\bibfnamefont {A.}~\bibnamefont {Macchi}},\
  }\href {\doibase 10.1103/PhysRevA.68.033201} {\bibfield  {journal} {\bibinfo
  {journal} {Phys. Rev. A}\ }\textbf {\bibinfo {volume} {68}},\ \bibinfo
  {pages} {033201} (\bibinfo {year} {2003})}\BibitemShut {NoStop}%
\bibitem [{\citenamefont {Ishikawa}\ and\ \citenamefont
  {Blenski}(2000)}]{Ishikawa}%
  \BibitemOpen
  \bibfield  {author} {\bibinfo {author} {\bibfnamefont {K.}~\bibnamefont
  {Ishikawa}}\ and\ \bibinfo {author} {\bibfnamefont {T.}~\bibnamefont
  {Blenski}},\ }\href {\doibase 10.1103/PhysRevA.62.063204} {\bibfield
  {journal} {\bibinfo  {journal} {Phys. Rev. A}\ }\textbf {\bibinfo {volume}
  {62}},\ \bibinfo {pages} {063204} (\bibinfo {year} {2000})}\BibitemShut
  {NoStop}%
\bibitem [{\citenamefont {Bauer}(2004)}]{Bauer2004}%
  \BibitemOpen
  \bibfield  {author} {\bibinfo {author} {\bibfnamefont {D.}~\bibnamefont
  {Bauer}},\ }\href {http://stacks.iop.org/0953-4075/37/i=15/a=007} {\bibfield
  {journal} {\bibinfo  {journal} {Journal of Physics B: Atomic, Molecular and
  Optical Physics}\ }\textbf {\bibinfo {volume} {37}},\ \bibinfo {pages} {3085}
  (\bibinfo {year} {2004})}\BibitemShut {NoStop}%
\bibitem [{\citenamefont {Megi}\ \emph {et~al.}(2003)\citenamefont {Megi},
  \citenamefont {Belkacem}, \citenamefont {Bouchene}, \citenamefont {Suraud},\
  and\ \citenamefont {Zwicknagel}}]{Megi}%
  \BibitemOpen
  \bibfield  {author} {\bibinfo {author} {\bibfnamefont {F.}~\bibnamefont
  {Megi}}, \bibinfo {author} {\bibfnamefont {M.}~\bibnamefont {Belkacem}},
  \bibinfo {author} {\bibfnamefont {M.~A.}\ \bibnamefont {Bouchene}}, \bibinfo
  {author} {\bibfnamefont {E.}~\bibnamefont {Suraud}}, \ and\ \bibinfo {author}
  {\bibfnamefont {G.}~\bibnamefont {Zwicknagel}},\ }\href
  {http://stacks.iop.org/0953-4075/36/i=2/a=308} {\bibfield  {journal}
  {\bibinfo  {journal} {Journal of Physics B: Atomic, Molecular and Optical
  Physics}\ }\textbf {\bibinfo {volume} {36}},\ \bibinfo {pages} {273}
  (\bibinfo {year} {2003})}\BibitemShut {NoStop}%
\bibitem [{\citenamefont {Kundu}\ and\ \citenamefont
  {Bauer}(2006{\natexlab{a}})}]{MKunduprl}%
  \BibitemOpen
  \bibfield  {author} {\bibinfo {author} {\bibfnamefont {M.}~\bibnamefont
  {Kundu}}\ and\ \bibinfo {author} {\bibfnamefont {D.}~\bibnamefont {Bauer}},\
  }\href {\doibase 10.1103/PhysRevLett.96.123401} {\bibfield  {journal}
  {\bibinfo  {journal} {Phys. Rev. Lett.}\ }\textbf {\bibinfo {volume} {96}},\
  \bibinfo {pages} {123401} (\bibinfo {year} {2006}{\natexlab{a}})}\BibitemShut
  {NoStop}%
\bibitem [{\citenamefont {Geindre}\ \emph {et~al.}(2010)\citenamefont
  {Geindre}, \citenamefont {Marjoribanks},\ and\ \citenamefont
  {Audebert}}]{Geindre_PRL2010}%
  \BibitemOpen
  \bibfield  {author} {\bibinfo {author} {\bibfnamefont {J.~P.}\ \bibnamefont
  {Geindre}}, \bibinfo {author} {\bibfnamefont {R.~S.}\ \bibnamefont
  {Marjoribanks}}, \ and\ \bibinfo {author} {\bibfnamefont {P.}~\bibnamefont
  {Audebert}},\ }\href {\doibase 10.1103/PhysRevLett.104.135001} {\bibfield
  {journal} {\bibinfo  {journal} {Phys. Rev. Lett.}\ }\textbf {\bibinfo
  {volume} {104}},\ \bibinfo {pages} {135001} (\bibinfo {year}
  {2010})}\BibitemShut {NoStop}%
\bibitem [{\citenamefont {Liseykina}\ \emph {et~al.}(2015)\citenamefont
  {Liseykina}, \citenamefont {Mulser},\ and\ \citenamefont
  {Murakami}}]{Mulser2015}%
  \BibitemOpen
  \bibfield  {author} {\bibinfo {author} {\bibfnamefont {T.}~\bibnamefont
  {Liseykina}}, \bibinfo {author} {\bibfnamefont {P.}~\bibnamefont {Mulser}}, \
  and\ \bibinfo {author} {\bibfnamefont {M.}~\bibnamefont {Murakami}},\ }\href
  {http://scitation.aip.org/content/aip/journal/pop/22/3/10.1063/1.4914837}
  {\bibfield  {journal} {\bibinfo  {journal} {Physics of Plasmas}\ }\textbf
  {\bibinfo {volume} {22}},\ \bibinfo {eid} {033302} (\bibinfo {year}
  {2015})}\BibitemShut {NoStop}%
\bibitem [{\citenamefont {Kundu}\ and\ \citenamefont
  {Bauer}(2006{\natexlab{b}})}]{MKundupra2006}%
  \BibitemOpen
  \bibfield  {author} {\bibinfo {author} {\bibfnamefont {M.}~\bibnamefont
  {Kundu}}\ and\ \bibinfo {author} {\bibfnamefont {D.}~\bibnamefont {Bauer}},\
  }\href {\doibase 10.1103/PhysRevA.74.063202} {\bibfield  {journal} {\bibinfo
  {journal} {Phys. Rev. A}\ }\textbf {\bibinfo {volume} {74}},\ \bibinfo
  {pages} {063202} (\bibinfo {year} {2006}{\natexlab{b}})}\BibitemShut
  {NoStop}%
\bibitem [{\citenamefont {Popruzhenko}\ \emph {et~al.}(2008)\citenamefont
  {Popruzhenko}, \citenamefont {Kundu}, \citenamefont {Zaretsky},\ and\
  \citenamefont {Bauer}}]{Popruzhenko2008}%
  \BibitemOpen
  \bibfield  {author} {\bibinfo {author} {\bibfnamefont {S.~V.}\ \bibnamefont
  {Popruzhenko}}, \bibinfo {author} {\bibfnamefont {M.}~\bibnamefont {Kundu}},
  \bibinfo {author} {\bibfnamefont {D.~F.}\ \bibnamefont {Zaretsky}}, \ and\
  \bibinfo {author} {\bibfnamefont {D.}~\bibnamefont {Bauer}},\ }\href
  {\doibase 10.1103/PhysRevA.77.063201} {\bibfield  {journal} {\bibinfo
  {journal} {Phys. Rev. A}\ }\textbf {\bibinfo {volume} {77}},\ \bibinfo
  {pages} {063201} (\bibinfo {year} {2008})}\BibitemShut {NoStop}%
\bibitem [{\citenamefont {Kundu}\ \emph {et~al.}(2012)\citenamefont {Kundu},
  \citenamefont {Kaw},\ and\ \citenamefont {Bauer}}]{MKundupra2012}%
  \BibitemOpen
  \bibfield  {author} {\bibinfo {author} {\bibfnamefont {M.}~\bibnamefont
  {Kundu}}, \bibinfo {author} {\bibfnamefont {P.~K.}\ \bibnamefont {Kaw}}, \
  and\ \bibinfo {author} {\bibfnamefont {D.}~\bibnamefont {Bauer}},\ }\href
  {\doibase 10.1103/PhysRevA.85.023202} {\bibfield  {journal} {\bibinfo
  {journal} {Phys. Rev. A}\ }\textbf {\bibinfo {volume} {85}},\ \bibinfo
  {pages} {023202} (\bibinfo {year} {2012})}\BibitemShut {NoStop}%
\bibitem [{\citenamefont {Mulser}\ \emph {et~al.}(2005)\citenamefont {Mulser},
  \citenamefont {Kanapathipillai},\ and\ \citenamefont {Hoffmann}}]{Mulserprl}%
  \BibitemOpen
  \bibfield  {author} {\bibinfo {author} {\bibfnamefont {P.}~\bibnamefont
  {Mulser}}, \bibinfo {author} {\bibfnamefont {M.}~\bibnamefont
  {Kanapathipillai}}, \ and\ \bibinfo {author} {\bibfnamefont {D.~H.~H.}\
  \bibnamefont {Hoffmann}},\ }\href {\doibase 10.1103/PhysRevLett.95.103401}
  {\bibfield  {journal} {\bibinfo  {journal} {Phys. Rev. Lett.}\ }\textbf
  {\bibinfo {volume} {95}},\ \bibinfo {pages} {103401} (\bibinfo {year}
  {2005})}\BibitemShut {NoStop}%
\bibitem [{\citenamefont {Mulser}\ and\ \citenamefont
  {Kanapathipillai}(2005)}]{Mulserpra}%
  \BibitemOpen
  \bibfield  {author} {\bibinfo {author} {\bibfnamefont {P.}~\bibnamefont
  {Mulser}}\ and\ \bibinfo {author} {\bibfnamefont {M.}~\bibnamefont
  {Kanapathipillai}},\ }\href {\doibase 10.1103/PhysRevA.71.063201} {\bibfield
  {journal} {\bibinfo  {journal} {Phys. Rev. A}\ }\textbf {\bibinfo {volume}
  {71}},\ \bibinfo {pages} {063201} (\bibinfo {year} {2005})}\BibitemShut
  {NoStop}%
\bibitem [{\citenamefont {Saalmann}\ and\ \citenamefont
  {Rost}(2003)}]{Saalmann2003}%
  \BibitemOpen
  \bibfield  {author} {\bibinfo {author} {\bibfnamefont {U.}~\bibnamefont
  {Saalmann}}\ and\ \bibinfo {author} {\bibfnamefont {J.-M.}\ \bibnamefont
  {Rost}},\ }\href {\doibase 10.1103/PhysRevLett.91.223401} {\bibfield
  {journal} {\bibinfo  {journal} {Phys. Rev. Lett.}\ }\textbf {\bibinfo
  {volume} {91}},\ \bibinfo {pages} {223401} (\bibinfo {year}
  {2003})}\BibitemShut {NoStop}%
\bibitem [{\citenamefont {Verlet}(1967)}]{Verlet}%
  \BibitemOpen
  \bibfield  {author} {\bibinfo {author} {\bibfnamefont {L.}~\bibnamefont
  {Verlet}},\ }\href {\doibase 10.1103/PhysRev.159.98} {\bibfield  {journal}
  {\bibinfo  {journal} {Phys. Rev.}\ }\textbf {\bibinfo {volume} {159}},\
  \bibinfo {pages} {98} (\bibinfo {year} {1967})}\BibitemShut {NoStop}%
\bibitem [{\citenamefont {Last}\ and\ \citenamefont
  {Jortner}(1999)}]{LastJortner1999}%
  \BibitemOpen
  \bibfield  {author} {\bibinfo {author} {\bibfnamefont {I.}~\bibnamefont
  {Last}}\ and\ \bibinfo {author} {\bibfnamefont {J.}~\bibnamefont {Jortner}},\
  }\href {\doibase 10.1103/PhysRevA.60.2215} {\bibfield  {journal} {\bibinfo
  {journal} {Phys. Rev. A}\ }\textbf {\bibinfo {volume} {60}},\ \bibinfo
  {pages} {2215} (\bibinfo {year} {1999})}\BibitemShut {NoStop}%
\bibitem [{\citenamefont {Last}\ and\ \citenamefont
  {Jortner}(2000)}]{LastJortner2000}%
  \BibitemOpen
  \bibfield  {author} {\bibinfo {author} {\bibfnamefont {I.}~\bibnamefont
  {Last}}\ and\ \bibinfo {author} {\bibfnamefont {J.}~\bibnamefont {Jortner}},\
  }\href {\doibase 10.1103/PhysRevA.62.013201} {\bibfield  {journal} {\bibinfo
  {journal} {Phys. Rev. A}\ }\textbf {\bibinfo {volume} {62}},\ \bibinfo
  {pages} {013201} (\bibinfo {year} {2000})}\BibitemShut {NoStop}%
\bibitem [{\citenamefont {Last}\ and\ \citenamefont
  {Jortner}(2004)}]{LastJortner2004}%
  \BibitemOpen
  \bibfield  {author} {\bibinfo {author} {\bibfnamefont {I.}~\bibnamefont
  {Last}}\ and\ \bibinfo {author} {\bibfnamefont {J.}~\bibnamefont {Jortner}},\
  }\href {\doibase http://dx.doi.org/10.1063/1.1630307} {\bibfield  {journal}
  {\bibinfo  {journal} {The Journal of Chemical Physics}\ }\textbf {\bibinfo
  {volume} {120}},\ \bibinfo {pages} {1336} (\bibinfo {year}
  {2004})}\BibitemShut {NoStop}%
\bibitem [{\citenamefont {Fomichev}\ \emph {et~al.}(2005)\citenamefont
  {Fomichev}, \citenamefont {Zaretsky}, \citenamefont {Bauer},\ and\
  \citenamefont {Becker}}]{Fomichev_pra}%
  \BibitemOpen
  \bibfield  {author} {\bibinfo {author} {\bibfnamefont {S.~V.}\ \bibnamefont
  {Fomichev}}, \bibinfo {author} {\bibfnamefont {D.~F.}\ \bibnamefont
  {Zaretsky}}, \bibinfo {author} {\bibfnamefont {D.}~\bibnamefont {Bauer}}, \
  and\ \bibinfo {author} {\bibfnamefont {W.}~\bibnamefont {Becker}},\ }\href
  {\doibase 10.1103/PhysRevA.71.013201} {\bibfield  {journal} {\bibinfo
  {journal} {Phys. Rev. A}\ }\textbf {\bibinfo {volume} {71}},\ \bibinfo
  {pages} {013201} (\bibinfo {year} {2005})}\BibitemShut {NoStop}%
\bibitem [{\citenamefont {Petrov}\ \emph
  {et~al.}(2005{\natexlab{a}})\citenamefont {Petrov}, \citenamefont {Davis},
  \citenamefont {Velikovich}, \citenamefont {Kepple}, \citenamefont {Dasgupta},
  \citenamefont {Clark}, \citenamefont {Borisov}, \citenamefont {Boyer},\ and\
  \citenamefont {Rhodes}}]{Petrov2005_PRE}%
  \BibitemOpen
  \bibfield  {author} {\bibinfo {author} {\bibfnamefont {G.~M.}\ \bibnamefont
  {Petrov}}, \bibinfo {author} {\bibfnamefont {J.}~\bibnamefont {Davis}},
  \bibinfo {author} {\bibfnamefont {A.~L.}\ \bibnamefont {Velikovich}},
  \bibinfo {author} {\bibfnamefont {P.~C.}\ \bibnamefont {Kepple}}, \bibinfo
  {author} {\bibfnamefont {A.}~\bibnamefont {Dasgupta}}, \bibinfo {author}
  {\bibfnamefont {R.~W.}\ \bibnamefont {Clark}}, \bibinfo {author}
  {\bibfnamefont {A.~B.}\ \bibnamefont {Borisov}}, \bibinfo {author}
  {\bibfnamefont {K.}~\bibnamefont {Boyer}}, \ and\ \bibinfo {author}
  {\bibfnamefont {C.~K.}\ \bibnamefont {Rhodes}},\ }\href {\doibase
  10.1103/PhysRevE.71.036411} {\bibfield  {journal} {\bibinfo  {journal} {Phys.
  Rev. E}\ }\textbf {\bibinfo {volume} {71}},\ \bibinfo {pages} {036411}
  (\bibinfo {year} {2005}{\natexlab{a}})}\BibitemShut {NoStop}%
\bibitem [{\citenamefont {Petrov}\ \emph
  {et~al.}(2005{\natexlab{b}})\citenamefont {Petrov}, \citenamefont {Davis},
  \citenamefont {Velikovich}, \citenamefont {Kepple}, \citenamefont
  {Dasgupta},\ and\ \citenamefont {Clark}}]{Petrov2005}%
  \BibitemOpen
  \bibfield  {author} {\bibinfo {author} {\bibfnamefont {G.~M.}\ \bibnamefont
  {Petrov}}, \bibinfo {author} {\bibfnamefont {J.}~\bibnamefont {Davis}},
  \bibinfo {author} {\bibfnamefont {A.~L.}\ \bibnamefont {Velikovich}},
  \bibinfo {author} {\bibfnamefont {P.}~\bibnamefont {Kepple}}, \bibinfo
  {author} {\bibfnamefont {A.}~\bibnamefont {Dasgupta}}, \ and\ \bibinfo
  {author} {\bibfnamefont {R.~W.}\ \bibnamefont {Clark}},\ }\href
  {http://scitation.aip.org/content/aip/journal/pop/12/6/10.1063/1.1928367}
  {\bibfield  {journal} {\bibinfo  {journal} {Physics of Plasmas}\ }\textbf
  {\bibinfo {volume} {12}},\ \bibinfo {eid} {063103} (\bibinfo {year}
  {2005}{\natexlab{b}})}\BibitemShut {NoStop}%
\bibitem [{\citenamefont {Petrov}\ and\ \citenamefont
  {Davis}(2006)}]{Petrov2006}%
  \BibitemOpen
  \bibfield  {author} {\bibinfo {author} {\bibfnamefont {G.~M.}\ \bibnamefont
  {Petrov}}\ and\ \bibinfo {author} {\bibfnamefont {J.}~\bibnamefont {Davis}},\
  }\href
  {http://scitation.aip.org/content/aip/journal/pop/13/3/10.1063/1.2167307}
  {\bibfield  {journal} {\bibinfo  {journal} {Physics of Plasmas}\ }\textbf
  {\bibinfo {volume} {13}},\ \bibinfo {eid} {033106} (\bibinfo {year}
  {2006})}\BibitemShut {NoStop}%
\bibitem [{\citenamefont {Davis}\ \emph {et~al.}(2007)\citenamefont {Davis},
  \citenamefont {Petrov},\ and\ \citenamefont {Velikovich}}]{Petrov2007}%
  \BibitemOpen
  \bibfield  {author} {\bibinfo {author} {\bibfnamefont {J.}~\bibnamefont
  {Davis}}, \bibinfo {author} {\bibfnamefont {G.~M.}\ \bibnamefont {Petrov}}, \
  and\ \bibinfo {author} {\bibfnamefont {A.}~\bibnamefont {Velikovich}},\
  }\href
  {http://scitation.aip.org/content/aip/journal/pop/14/6/10.1063/1.2743646}
  {\bibfield  {journal} {\bibinfo  {journal} {Physics of Plasmas}\ }\textbf
  {\bibinfo {volume} {14}},\ \bibinfo {eid} {060701} (\bibinfo {year}
  {2007})}\BibitemShut {NoStop}%
\bibitem [{\citenamefont {Mishra}\ \emph {et~al.}(2011)\citenamefont {Mishra},
  \citenamefont {Holkundkar},\ and\ \citenamefont {Gupta}}]{G_Mishra2011}%
  \BibitemOpen
  \bibfield  {author} {\bibinfo {author} {\bibfnamefont {G.}~\bibnamefont
  {Mishra}}, \bibinfo {author} {\bibfnamefont {A.~R.}\ \bibnamefont
  {Holkundkar}}, \ and\ \bibinfo {author} {\bibfnamefont {N.}~\bibnamefont
  {Gupta}},\ }\href {\doibase 10.1017/S0263034611000346} {\bibfield  {journal}
  {\bibinfo  {journal} {Laser and Particle Beams}\ }\textbf {\bibinfo {volume}
  {29}},\ \bibinfo {pages} {305} (\bibinfo {year} {2011})}\BibitemShut
  {NoStop}%
\bibitem [{\citenamefont {Mishra}\ and\ \citenamefont
  {Gupta}(2012)}]{G_Mishra}%
  \BibitemOpen
  \bibfield  {author} {\bibinfo {author} {\bibfnamefont {G.}~\bibnamefont
  {Mishra}}\ and\ \bibinfo {author} {\bibfnamefont {N.~K.}\ \bibnamefont
  {Gupta}},\ }\href
  {http://scitation.aip.org/content/aip/journal/pop/19/9/10.1063/1.4752016}
  {\bibfield  {journal} {\bibinfo  {journal} {Physics of Plasmas}\ }\textbf
  {\bibinfo {volume} {19}},\ \bibinfo {eid} {093107} (\bibinfo {year}
  {2012})}\BibitemShut {NoStop}%
\bibitem [{\citenamefont {Cheng}\ \emph {et~al.}(2015)\citenamefont {Cheng},
  \citenamefont {Zhang}, \citenamefont {Fu},\ and\ \citenamefont
  {Liu}}]{Cheng}%
  \BibitemOpen
  \bibfield  {author} {\bibinfo {author} {\bibfnamefont {R.}~\bibnamefont
  {Cheng}}, \bibinfo {author} {\bibfnamefont {C.}~\bibnamefont {Zhang}},
  \bibinfo {author} {\bibfnamefont {L.-B.}\ \bibnamefont {Fu}}, \ and\ \bibinfo
  {author} {\bibfnamefont {J.}~\bibnamefont {Liu}},\ }\href
  {http://stacks.iop.org/0953-4075/48/i=3/a=035601} {\bibfield  {journal}
  {\bibinfo  {journal} {Journal of Physics B: Atomic, Molecular and Optical
  Physics}\ }\textbf {\bibinfo {volume} {48}},\ \bibinfo {pages} {035601}
  (\bibinfo {year} {2015})}\BibitemShut {NoStop}%
\bibitem [{\citenamefont {Holkundkar}\ \emph {et~al.}(2011)\citenamefont
  {Holkundkar}, \citenamefont {Mishra},\ and\ \citenamefont {Gupta}}]{Amol}%
  \BibitemOpen
  \bibfield  {author} {\bibinfo {author} {\bibfnamefont {A.~R.}\ \bibnamefont
  {Holkundkar}}, \bibinfo {author} {\bibfnamefont {G.}~\bibnamefont {Mishra}},
  \ and\ \bibinfo {author} {\bibfnamefont {N.~K.}\ \bibnamefont {Gupta}},\
  }\href
  {http://scitation.aip.org/content/aip/journal/pop/18/5/10.1063/1.3581061}
  {\bibfield  {journal} {\bibinfo  {journal} {Physics of Plasmas}\ }\textbf
  {\bibinfo {volume} {18}},\ \bibinfo {eid} {053102} (\bibinfo {year}
  {2011})}\BibitemShut {NoStop}%
\bibitem [{\citenamefont {Greschik}\ \emph {et~al.}(2005)\citenamefont
  {Greschik}, \citenamefont {Arndt},\ and\ \citenamefont
  {Kull}}]{Greschik&Kull}%
  \BibitemOpen
  \bibfield  {author} {\bibinfo {author} {\bibfnamefont {F.}~\bibnamefont
  {Greschik}}, \bibinfo {author} {\bibfnamefont {L.}~\bibnamefont {Arndt}}, \
  and\ \bibinfo {author} {\bibfnamefont {H.-J.}\ \bibnamefont {Kull}},\ }\href
  {http://stacks.iop.org/0295-5075/72/i=3/a=376} {\bibfield  {journal}
  {\bibinfo  {journal} {EPL (Europhysics Letters)}\ }\textbf {\bibinfo {volume}
  {72}},\ \bibinfo {pages} {376} (\bibinfo {year} {2005})}\BibitemShut
  {NoStop}%
\bibitem [{\citenamefont {Batishchev}\ \emph {et~al.}(2004)\citenamefont
  {Batishchev}, \citenamefont {Batishcheva}, \citenamefont {Bychenkov},
  \citenamefont {Albukrek}, \citenamefont {Brantov},\ and\ \citenamefont
  {Rozmus}}]{Batishchev}%
  \BibitemOpen
  \bibfield  {author} {\bibinfo {author} {\bibfnamefont {O.}~\bibnamefont
  {Batishchev}}, \bibinfo {author} {\bibfnamefont {A.}~\bibnamefont
  {Batishcheva}}, \bibinfo {author} {\bibfnamefont {V.}~\bibnamefont
  {Bychenkov}}, \bibinfo {author} {\bibfnamefont {C.}~\bibnamefont {Albukrek}},
  \bibinfo {author} {\bibfnamefont {A.}~\bibnamefont {Brantov}}, \ and\
  \bibinfo {author} {\bibfnamefont {W.}~\bibnamefont {Rozmus}},\ }\href
  {\doibase 10.1016/j.cpc.2004.06.080} {\bibfield  {journal} {\bibinfo
  {journal} {Computer Physics Communications}\ }\textbf {\bibinfo {volume}
  {164}},\ \bibinfo {pages} {53 } (\bibinfo {year} {2004})}\BibitemShut
  {NoStop}%
\bibitem [{\citenamefont {Bystryi}\ and\ \citenamefont
  {Morozov}(2015)}]{Bystryi}%
  \BibitemOpen
  \bibfield  {author} {\bibinfo {author} {\bibfnamefont {R.~G.}\ \bibnamefont
  {Bystryi}}\ and\ \bibinfo {author} {\bibfnamefont {I.~V.}\ \bibnamefont
  {Morozov}},\ }\href {http://stacks.iop.org/0953-4075/48/i=1/a=015401}
  {\bibfield  {journal} {\bibinfo  {journal} {Journal of Physics B: Atomic,
  Molecular and Optical Physics}\ }\textbf {\bibinfo {volume} {48}},\ \bibinfo
  {pages} {015401} (\bibinfo {year} {2015})}\BibitemShut {NoStop}%
\bibitem [{\citenamefont {Arbeiter}\ and\ \citenamefont
  {Fennel}(2010)}]{Arbeiter}%
  \BibitemOpen
  \bibfield  {author} {\bibinfo {author} {\bibfnamefont {M.}~\bibnamefont
  {Arbeiter}}\ and\ \bibinfo {author} {\bibfnamefont {T.}~\bibnamefont
  {Fennel}},\ }\href {\doibase 10.1103/PhysRevA.82.013201} {\bibfield
  {journal} {\bibinfo  {journal} {Phys. Rev. A}\ }\textbf {\bibinfo {volume}
  {82}},\ \bibinfo {pages} {013201} (\bibinfo {year} {2010})}\BibitemShut
  {NoStop}%
\bibitem [{\citenamefont {D\"oppner}\ \emph {et~al.}(2005)\citenamefont
  {D\"oppner}, \citenamefont {Fennel}, \citenamefont {Diederich}, \citenamefont
  {Tiggesb\"aumker},\ and\ \citenamefont {Meiwes-Broer}}]{Doppner}%
  \BibitemOpen
  \bibfield  {author} {\bibinfo {author} {\bibfnamefont {T.}~\bibnamefont
  {D\"oppner}}, \bibinfo {author} {\bibfnamefont {T.}~\bibnamefont {Fennel}},
  \bibinfo {author} {\bibfnamefont {T.}~\bibnamefont {Diederich}}, \bibinfo
  {author} {\bibfnamefont {J.}~\bibnamefont {Tiggesb\"aumker}}, \ and\ \bibinfo
  {author} {\bibfnamefont {K.~H.}\ \bibnamefont {Meiwes-Broer}},\ }\href
  {\doibase 10.1103/PhysRevLett.94.013401} {\bibfield  {journal} {\bibinfo
  {journal} {Phys. Rev. Lett.}\ }\textbf {\bibinfo {volume} {94}},\ \bibinfo
  {pages} {013401} (\bibinfo {year} {2005})}\BibitemShut {NoStop}%
\bibitem [{\citenamefont {K\"oller}\ \emph {et~al.}(1999)\citenamefont
  {K\"oller}, \citenamefont {Schumacher}, \citenamefont {K\"ohn}, \citenamefont
  {Teuber}, \citenamefont {Tiggesb\"aumker},\ and\ \citenamefont
  {Meiwes-Broer}}]{Koller}%
  \BibitemOpen
  \bibfield  {author} {\bibinfo {author} {\bibfnamefont {L.}~\bibnamefont
  {K\"oller}}, \bibinfo {author} {\bibfnamefont {M.}~\bibnamefont
  {Schumacher}}, \bibinfo {author} {\bibfnamefont {J.}~\bibnamefont {K\"ohn}},
  \bibinfo {author} {\bibfnamefont {S.}~\bibnamefont {Teuber}}, \bibinfo
  {author} {\bibfnamefont {J.}~\bibnamefont {Tiggesb\"aumker}}, \ and\ \bibinfo
  {author} {\bibfnamefont {K.~H.}\ \bibnamefont {Meiwes-Broer}},\ }\href
  {\doibase 10.1103/PhysRevLett.82.3783} {\bibfield  {journal} {\bibinfo
  {journal} {Phys. Rev. Lett.}\ }\textbf {\bibinfo {volume} {82}},\ \bibinfo
  {pages} {3783} (\bibinfo {year} {1999})}\BibitemShut {NoStop}%
\bibitem [{\citenamefont {Zamith}\ \emph {et~al.}(2004)\citenamefont {Zamith},
  \citenamefont {Martchenko}, \citenamefont {Ni}, \citenamefont {Aseyev},
  \citenamefont {Muller},\ and\ \citenamefont {Vrakking}}]{Zamith}%
  \BibitemOpen
  \bibfield  {author} {\bibinfo {author} {\bibfnamefont {S.}~\bibnamefont
  {Zamith}}, \bibinfo {author} {\bibfnamefont {T.}~\bibnamefont {Martchenko}},
  \bibinfo {author} {\bibfnamefont {Y.}~\bibnamefont {Ni}}, \bibinfo {author}
  {\bibfnamefont {S.~A.}\ \bibnamefont {Aseyev}}, \bibinfo {author}
  {\bibfnamefont {H.~G.}\ \bibnamefont {Muller}}, \ and\ \bibinfo {author}
  {\bibfnamefont {M.~J.~J.}\ \bibnamefont {Vrakking}},\ }\href {\doibase
  10.1103/PhysRevA.70.011201} {\bibfield  {journal} {\bibinfo  {journal} {Phys.
  Rev. A}\ }\textbf {\bibinfo {volume} {70}},\ \bibinfo {pages} {011201}
  (\bibinfo {year} {2004})}\BibitemShut {NoStop}%
\bibitem [{\citenamefont {Taguchi}\ \emph {et~al.}(2004)\citenamefont
  {Taguchi}, \citenamefont {Antonsen},\ and\ \citenamefont
  {Milchberg}}]{Taguchi_PRl}%
  \BibitemOpen
  \bibfield  {author} {\bibinfo {author} {\bibfnamefont {T.}~\bibnamefont
  {Taguchi}}, \bibinfo {author} {\bibfnamefont {T.~M.}\ \bibnamefont
  {Antonsen}}, \ and\ \bibinfo {author} {\bibfnamefont {H.~M.}\ \bibnamefont
  {Milchberg}},\ }\href {\doibase 10.1103/PhysRevLett.92.205003} {\bibfield
  {journal} {\bibinfo  {journal} {Phys. Rev. Lett.}\ }\textbf {\bibinfo
  {volume} {92}},\ \bibinfo {pages} {205003} (\bibinfo {year}
  {2004})}\BibitemShut {NoStop}%
\bibitem [{\citenamefont {Antonsen}\ \emph {et~al.}(2005)\citenamefont
  {Antonsen}, \citenamefont {Taguchi}, \citenamefont {Gupta}, \citenamefont
  {Palastro},\ and\ \citenamefont {Milchberg}}]{Antonsen}%
  \BibitemOpen
  \bibfield  {author} {\bibinfo {author} {\bibfnamefont {T.~M.}\ \bibnamefont
  {Antonsen}}, \bibinfo {author} {\bibfnamefont {T.}~\bibnamefont {Taguchi}},
  \bibinfo {author} {\bibfnamefont {A.}~\bibnamefont {Gupta}}, \bibinfo
  {author} {\bibfnamefont {J.}~\bibnamefont {Palastro}}, \ and\ \bibinfo
  {author} {\bibfnamefont {H.~M.}\ \bibnamefont {Milchberg}},\ }\href
  {http://scitation.aip.org/content/aip/journal/pop/12/5/10.1063/1.1869500}
  {\bibfield  {journal} {\bibinfo  {journal} {Physics of Plasmas}\ }\textbf
  {\bibinfo {volume} {12}},\ \bibinfo {eid} {056703} (\bibinfo {year}
  {2005})}\BibitemShut {NoStop}%
\bibitem [{\citenamefont {Kostyukov}\ and\ \citenamefont
  {Rax}(2003)}]{Kostyukov}%
  \BibitemOpen
  \bibfield  {author} {\bibinfo {author} {\bibfnamefont {I.}~\bibnamefont
  {Kostyukov}}\ and\ \bibinfo {author} {\bibfnamefont {J.-M.}\ \bibnamefont
  {Rax}},\ }\href {\doibase 10.1103/PhysRevE.67.066405} {\bibfield  {journal}
  {\bibinfo  {journal} {Phys. Rev. E}\ }\textbf {\bibinfo {volume} {67}},\
  \bibinfo {pages} {066405} (\bibinfo {year} {2003})}\BibitemShut {NoStop}%
\end{thebibliography}%

\end{document}